\documentclass[english,aps,prb,amsmath,amssymb,twocolumn,letterpaper,showpacs,superscriptaddress]{revtex4-2}

\usepackage{mathtools} 

\usepackage[normalem]{ulem} 

\usepackage{comment}
\usepackage{amsmath}
\usepackage{physics}
\usepackage{amssymb}
\usepackage{pifont}
\usepackage{graphicx,bm}
\usepackage{makecell,multirow}
\usepackage{babel}
\usepackage{amstext}
\usepackage{esint}
\usepackage{cases}
\usepackage{makecell}
\usepackage{lineno}

\usepackage{xcolor}
\definecolor{cornflowerblue}{HTML}{80549A}
\usepackage[caption=false]{subfig}
\usepackage{verbatim}
\usepackage[unicode=true,pdfusetitle,
bookmarks=true,bookmarksnumbered=false,bookmarksopen=false, breaklinks=false,pdfborder={0 0 1},backref=false,colorlinks=false]
 {hyperref}
 \hypersetup{
    colorlinks=true,
    linkcolor=red,
    citecolor=blue,
    filecolor=magenta,      
    urlcolor=blue
    }
\usepackage{pdfpages}	
\usepackage{pgffor}		
\usepackage{upgreek}
\usepackage{braket}

\makeatletter
\AtBeginDocument{\let\LS@rot\@undefined} 
\makeatother							 

\begin{document}

\title{Krylov complexity and fidelity susceptibility in two-band Hamiltonians}

\author{Rishav Chaudhuri}
\thanks{These authors contributed equally to this work.}
\affiliation{Department of Physics and Astronomy, Purdue University, West Lafayette, Indiana 47907, USA}
\affiliation{Purdue Quantum Science and Engineering Institute, West Lafayette, Indiana 47907, USA}

\author{Ayush Raj}
\thanks{These authors contributed equally to this work.}
\affiliation{Department of Physics and Astronomy, Purdue University, West Lafayette, Indiana 47907, USA}

\author{Soham Ray}
\thanks{These authors contributed equally to this work.}
\affiliation{Department of Physics and Astronomy, Purdue University, West Lafayette, Indiana 47907, USA}

\author{Sai Satyam Samal}
\thanks{These authors contributed equally to this work.}
\affiliation{Department of Physics and Astronomy, Purdue University, West Lafayette, Indiana 47907, USA}

\newcommand*{\RCc}[1]{\textcolor{green}{[RC: #1]}}
\newcommand{\RC}[1]{\textcolor{green}{#1}}

\begin{abstract}

We investigate Krylov spread complexity for the ground state of two-band Hamiltonians, where the reference state is a generic state on the Bloch sphere. The spread complexity is obtained by using a purely geometric formulation in terms of Bloch sphere data without constructing the circuit Hamiltonian. For generic reference states, the derivative of the spread complexity is logarithmically divergent at the topological phase transition in the Su-Schrieffer-Heeger (SSH) model. We demonstrate that the derivative of the spread complexity is bounded by fidelity susceptibility for general two-band models, indicating the sensitivity of the spread complexity to any gap closing (topological or trivial). This is illustrated in the massive Dirac Hamiltonian with a trivial gap closing. Finally, we introduce a non-unitary duality in the SSH model between the topological and trivial phases, which manifests itself in the spread complexity and fidelity susceptibility.

\end{abstract}

\maketitle

\begingroup
\hypersetup{linkcolor=cornflowerblue}
\tableofcontents
\endgroup

\section{Introduction} Understanding how quantum states reorganize across phase transitions is a central theme in condensed matter physics~\cite{landau1980statistical,RevModPhys.69.315,Osborne2002,doi:10.1126/science.1121541,10.1093/acprof:oso/9780199227259.001.0001,Sachdev2011,PhysRevX.5.031007,PhysRevB.100.134203,PhysRevB.100.134306,PhysRevResearch.2.013323,PhysRevB.101.174305,ALI2020135919,10.21468/SciPostPhys.15.5.186,annurev:/content/journals/10.1146/annurev-conmatphys-031720-030658,sym15051055}. These reorganizations are commonly probed through order parameters~\cite{Kardar_2007}, topological invariants~\cite{Alicea_2012,6ee78b01-9610-3faf-a626-578be4e5d3b4,Asboth2016}, and fidelity susceptibility ~\cite{PhysRevA.76.062318, PhysRevA.78.010301,PhysRevA.79.032302,PhysRevB.102.134111,PhysRevA.111.032401}, which characterize symmetry breaking, underlying topology, or changes in the ground state geometry. Quantum complexity~\cite{JeffersonMyers2017QFTComplexity,HacklMyers2018Fermions,Haferkamp2022,PhysRevD.106.046007,PhysRevD.110.084039,doi:10.1073/pnas.2415913122,10.21468/SciPostPhys.15.5.186} offers a complementary perspective by quantifying the difficulty of preparing a target state from a chosen reference state. This viewpoint has motivated recent work on Krylov spread complexity~\cite{Caputa:2021sib,Bhattacharya:2022gbz,Bhattacharya:2023zqt,Bhattacharjee:2023uwx, PhysRevD.109.046017, caputa_2024, Nandy:2024evd,Chakrabarti_2025, Balasubramanian2026,Cafaro2026KrylovsSC,pain2026krylovspaceanatomyspreadcomplexity,grabarits2026krylovdynamicsoperatorgrowth} as a diagnostic of phases of matter, suggesting that the structure of state preparation may encode signatures of criticality and topology~\cite{Camargo:2018eof,PhysRevB.106.195125,Caputa2023,PhysRevB.111.174207,6lvg-7qdn}.

In practice, evaluating the Krylov spread complexity of ground states in generic quantum systems possesses a significant technical obstruction. This is due to the requirement of a circuit Hamiltonian that connects the reference state to the target ground state. Such a Hamiltonian is generally not unique and is difficult to construct~\cite{PhysRevB.106.195125,Caputa2023,PhysRevB.111.174207,6lvg-7qdn}. In addition, the spread complexity depends explicitly on the choice of reference state, making it a potentially ambiguous probe of phase transitions. In particular, topological phase transitions are accompanied by a gap closing~\cite{10.1093/acprof:oso/9780199227259.001.0001,6ee78b01-9610-3faf-a626-578be4e5d3b4,Alicea_2012,Asboth2016}, however, not all gap closings involve a change of topology~\cite{Sachdev2011}. It is therefore important to determine whether spread complexity detects only topological phase transitions, or if it is sensitive to all gap closings.

In this article, we address this ambiguity by studying generic two-band Hamiltonians. We obtain a geometric expression for the ground state Krylov spread complexity in terms of Bloch sphere data, without explicitly constructing the circuit Hamiltonian. This formulation makes the reference state dependence of the complexity evident. For generic reference states, the derivative of the spread complexity becomes singular at gap closings. We relate this response to fidelity susceptibility through a general bound, establishing the derivative of the complexity as a probe of gap closings rather than topology alone. 
We illustrate these results in two settings, the Su-Schrieffer-Heeger (SSH) model~\cite{PhysRevLett.42.1698,Su1980,JACKIW1981253}, where the gap closing accompanies a topological phase transition, and a topologically trivial massive Dirac model. We further show that a non-unitary duality in the SSH model constrains both the complexity and fidelity susceptibility across the two phases.

\section{Krylov spread complexity for two-band models} We begin with a reference state $\ket{\psi_{\rm ref}}$ and choose a circuit Hamiltonian $H_{\rm C}$ such that the unitary operator $e^{-iH_{\rm C}t}$ maps $\ket{\psi_{\rm ref}}$ to the target state $\ket{\psi_{\rm target}}$ in unit time, i.e., $e^{-iH_{\rm C}}\ket{\psi_{\rm ref}}=\ket{\psi_{\rm target}}$~\cite{PhysRevB.106.195125,6lvg-7qdn}.
The Krylov spread complexity is a measure of the distance between $\ket{\psi_{\text{target}}}$ and $\ket{\psi_{\text{ref}}}$ in the Hilbert space. This is quantified by expanding $\ket{\psi_{\text{target}}}$ in terms of orthogonal basis vectors called the Krylov basis states. The Krylov basis states are obtained by a Gram-Schmidt orthogonalization~\cite{Lanczos1950,PhysRevX.9.041017} of states $\{ \ket{\psi_{\text{ref}}}, H_{\text{C}} \ket{\psi_{\text{ref}}}, H_{\text{C}}^2 \ket{\psi_{\text{ref}}}, \cdots \}$. We denote the Krylov basis states as $\{ \ket{\mathcal{K}_0}, \ket{\mathcal{K}_1}, \ket{\mathcal{K}_2}, \cdots\}$ where $\ket{\mathcal{K}_0} = \ket{\psi_{\text{ref}}}$. We now expand the target state in the Krylov basis as $\ket{\psi_{\text{target}}} = \sum_n \psi_n \ket{\mathcal{K}_n}$ where $\sum_{n}|\psi_{n}|^{2} = 1$. Finally, the Krylov spread complexity of $\ket{\psi_{\text{target}}}$ with respect to $\ket{\psi_{\text{ref}}}$ and with the circuit Hamiltonian $H_{\text{C}}$ is,
    \begin{equation}\label{eq:krylov_spread_complexity}
        C = \sum_n n |\psi_n|^2\,.
    \end{equation}

    In general, given a reference state $\ket{\psi_{\text{ref}}}$ and a target state $\ket{\psi_{\text{target}}}$, it is a difficult problem to find the circuit Hamiltonian $H_{\text{C}}$ such that $e^{-iH_c}\ket{\psi_{\text{ref}}}=\ket{\psi_{\text{target}}}$. However, for a two-band Hamiltonian, the spread complexity is independent of the choice of $H_{\text{C}}$~\cite{PhysRevD.109.L081701,PhysRevD.111.076014,Cafaro2026KrylovsSC}. We work with a two-band Hamiltonian, $H= \sum_k \bold{d}(k)\cdot\boldsymbol{\sigma}$ where $\bold{d}(k)=[d_{x}(k),\,d_{y}(k),\,d_{z}(k)]$ encapsulates all the details of the system and $\boldsymbol{\sigma}=(\sigma_{x},\,\sigma_{y},\,\sigma_{z})$ are the Pauli matrices. We choose the ground state of the Hamiltonian as our target state, $\ket{\psi_{\text{target}}} = \otimes_k \ket{\psi_{\text{target}}}_k$, where $\bold{d}(k)\cdot\boldsymbol{\sigma} \ket{\psi_{\text{target}}}_k = -|\bold{d}(k)| \cdot \ket{\psi_{\text{target}}}_k$ and $|\bold{d}(k)|=\sqrt{d_{x}^{2}(k)+d_{y}^{2}(k)+d_{z}^{2}(k)}$. The Bloch sphere vector corresponding to the ground state $\ket{\psi_{\text{target}}}_k$ is $-\left [ d_x(k),\,d_y(k),\,d_z(k) \right ]/|\bold{d}(k)|$.

    Next, we choose a reference state
$\ket{\psi_{\rm ref}}=\otimes_k\ket{\psi_{\rm ref}}_k$, where 
$\ket{\psi_{\rm ref}}_k=\alpha\ket{\uparrow}_k+\beta\ket{\downarrow}_k$ is a momentum-independent vector on the Bloch sphere,
with $\alpha,\beta\in\mathbb{C}$ satisfying
$|\alpha|^2+|\beta|^2=1$. From here on, we drop the $k$ label on $\ket{\uparrow}_k$ and $\ket{\downarrow}_k$ for clarity. The states $\ket{\uparrow}$ and $\ket{\downarrow}$ are the eigenstates of $\sigma_z$ corresponding to the eigenvalues $\pm 1$ respectively. On the Bloch sphere, $\ket{\psi_{\text{ref}}}_k$ corresponds to the vector $(\sin \theta \cos \phi,\, \sin \theta \sin \phi ,\, \cos \theta)$, where $\alpha=\cos \left ( \frac{\theta}{2} \right )$ and $\beta = e^{i\phi} \sin \left ( \frac{\theta}{2} \right )$. The direct product structure of the reference and the target state allows us to find a circuit Hamiltonian $H_{\text{C}}(k)$ for each $k$ such that $e^{-iH_{\text{C}}(k)}\ket{\psi_{\text{ref}}}_{k} = \ket{\psi_{\text{target}}}_{k}$, and hence, we have $e^{-iH_{\text{C}}} = \otimes_k e^{-iH_{\text{C}}(k)}$. For each $k$, the Hilbert space is $2-$dimensional and thus the Krylov subspace is also $2-$dimensional. With $\ket{\mathcal{K}_0}_k = \ket{\psi_{\text{ref}}}_k$, we now identify the second Krylov basis vector $\ket{\mathcal{K}_1}_k$ as $\ket{\psi_{\text{ref}}^\perp}_k$ up to a phase such that ${}_k\!\braket{\psi_{\text{ref}}^\perp|\psi_{\text{ref}}}_{k}=0$. Thus, the explicit form of $H_{\text{C}}$ is not required for determining the full Krylov basis for two-band models.

     Expanding the target state in the Krylov basis, we have $\ket{\psi_{\text{target}}}_k = \psi_0 \ket{\psi_{\text{ref}}}_k + \psi_1 \ket{\psi_{\text{ref}}^\perp}_k$, with $|\psi_{0}|^{2} + |\psi_{1}|^{2} = 1$. By using Eq.~(\ref{eq:krylov_spread_complexity}), we find $C_k  = 1 - |\psi_{0}|^{2} = 1-|{}_k\!\braket{\psi_{\text{ref}}|\psi_{\text{target}}}_k|^2$. Next, by using Bloch sphere geometry, we find $|{}_k\!\braket{\psi_{\text{ref}}|\psi_{\text{target}}}_k|^2 = [1+\hat{n}_{\text{ref}}(k) \cdot \hat{n}_{\text{target}}(k)]/2$, where $\hat{n}_{\text{ref}}(k)$ and $\hat{n}_{\text{target}}(k)$ are the Bloch sphere vectors corresponding to $\ket{\psi_{\text{ref}}}_k$ and $\ket{\psi_{\text{target}}}_k$ respectively. Hence, the Krylov spread complexity for a given momentum $k$ is given as follows (see Appendix~\ref{appendix:b}),
    \begin{equation}\label{eq:ck_overlap_bloch}
        C_k = \frac{1-\hat{n}_{\text{ref}}(k) \cdot \hat{n}_{\text{target}}(k)}{2}\,.
    \end{equation}
    From here we obtain the average spread complexity over the full Brillouin zone, 
    \begin{align}\label{eq:ck_integral}
        C = \frac{1}{2\pi}\int_{-\pi}^{\pi} C_k dk\,.
    \end{align}

    \section{Su-Schrieffer-Heeger model} 
\begin{figure}[t!]
    \centering
    \includegraphics[width=1.0\linewidth]{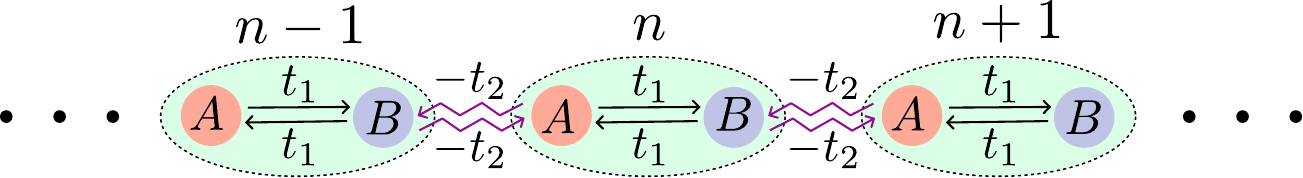}
    \caption{Schematic of the Su-Schrieffer-Heeger (SSH) model. Each lattice site contains two species of fermions, $A$ and $B$, with $t_{1}$ as the intercell hopping amplitude and $-t_2$ as the intracell hopping amplitude.}
    \label{fig:ssh_schematic}
    \end{figure}

We now illustrate the preceding formalism in the Su-Schrieffer-Heeger (SSH) model~\cite{PhysRevLett.42.1698,Su1980,JACKIW1981253}, which exhibits a topological phase transition. The SSH Hamiltonian is given by,
    \begin{equation}\label{eq:ssh_hamiltonian}
        H_{\text{SSH}} = \sum_{n=1}^{L}\left( t_{1}c_{A,n}^{\dagger}c_{B,n} -t_{2}c_{B,n}^{\dagger}c_{A,n+1} + h.c. \right) ,
    \end{equation}
    with intracell hopping $t_1$ and intercell hopping $-t_2$, see Fig.~\ref{fig:ssh_schematic}. The Hamiltonian in the momentum space after a unitary transformation $(\sigma_x,\sigma_y,\sigma_z) \to (\sigma_x,\sigma_z,-\sigma_y)$ is given as $H_{\text{SSH}}(k) = \bold{d}(k)\cdot \boldsymbol{\sigma} = d_{x}(k)\sigma_{x} + d_{z}(k) \sigma_{z}$ where $d_{x}(k) = t_{1} - t_{2}\cos k$, and $d_{z}(k) = t_{2}\sin k$, see Appendix~\ref{appendix:a}. The Bloch sphere vector corresponding to the ground state or equivalently, the target state in our case is given as follows,
    \begin{equation}\label{eq:boch_ssh_ground_state}
        \hat{d}_{\text{SSH}}(k)=\frac{(t_1 - t_2 \cos k,\, 0,\, t_2 \sin k)}{|\bold{d}_{\text{SSH}}(k)|} = -\hat{n}_{\text{target}}(k) ,
    \end{equation}
    where $|\bold{d}_{\text{SSH}}(k)|=\sqrt{t_1^2 + t_2^2 - 2t_1 t_2 \cos k}$.
     
\begin{figure*}[htbp]
    \centering
    \includegraphics[width=1.0\linewidth]{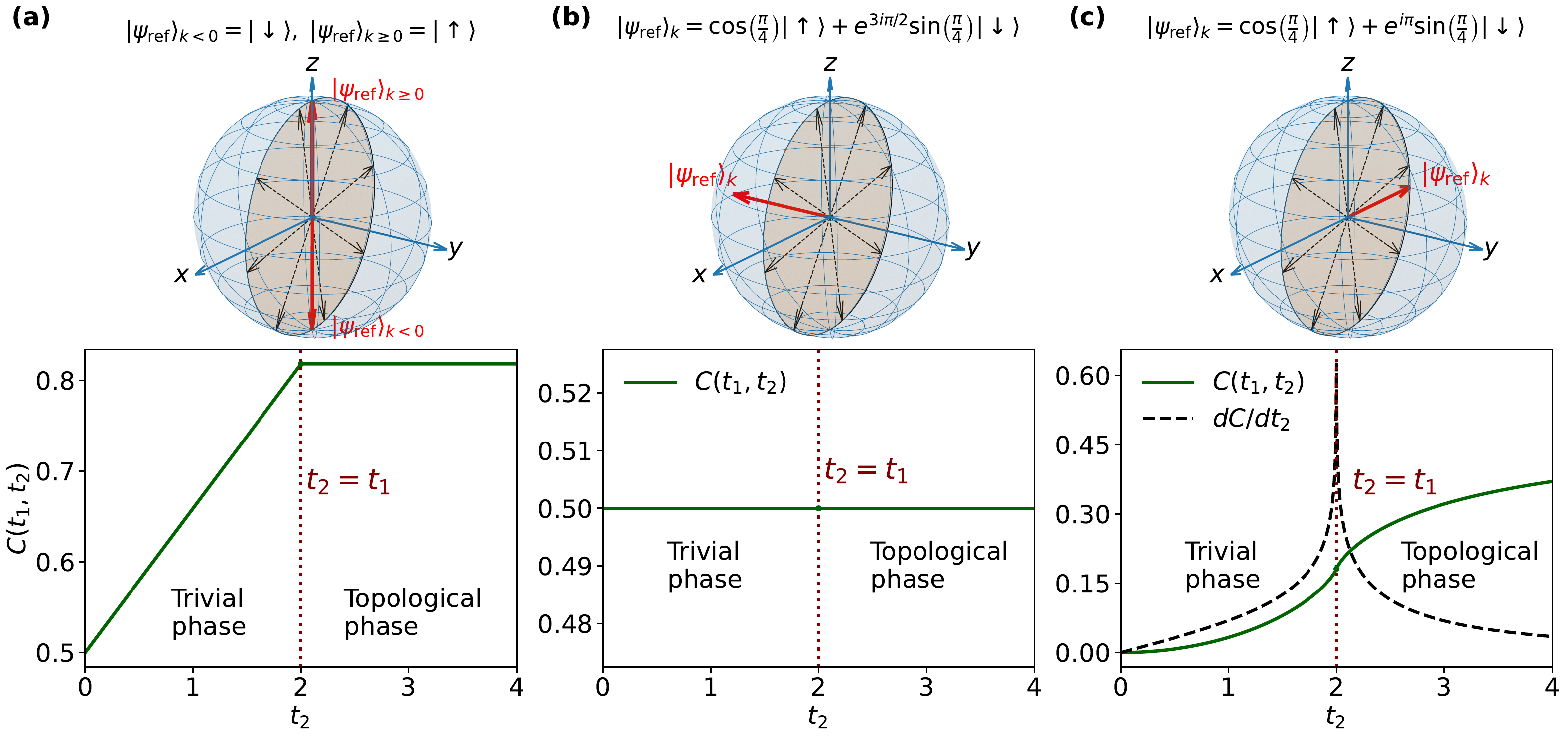}
    \caption{Krylov spread complexity for the ground state of the Su-Schrieffer-Heeger (SSH) model for different choices of reference state. The shaded region with dashed arrows along the $x-z$ plane represents the ground state subspace of the SSH Hamiltonian, $H_{\text{SSH}}(k) = \bold{d}(k)\cdot \boldsymbol{\sigma} = d_{x}(k)\sigma_{x} + d_{z}(k) \sigma_{z}$ where $d_{x}(k) = t_{1} - t_{2}\cos k$, and $d_{z}(k) = t_{2}\sin k$. (a) The reference state $\ket{\psi_{\text{ref}}}$ is momentum-dependent, and the corresponding Bloch sphere vector is $\hat{n}_{\text{ref}}(k)=(0,0,-1)$ for $k<0$ and $\hat{n}_{\text{ref}}(k)=(0,0,1)$ for $k \geq 0$. The spread complexity is linearly dependent on $t_2$ in the trivial phase and shows a constant plateau in the topological phase, recovering earlier results~\cite{PhysRevB.106.195125}. (b) The spread complexity is a constant across both the trivial and the topological phase for $\hat{n}_{\text{ref}}(k)=(0,-1,0)$. (c) With $\hat{n}_{\text{ref}}(k)=( -1,0,0)$, the derivative of the spread complexity with respect to $t_2$ shows a logarithmic divergence at the topological phase transition point $t_1 = t_2$.}
    \label{fig:bloch_spheres_complexity_plots}
    \end{figure*}

    Using Eq.~(\ref{eq:ck_overlap_bloch}), with reference state $\ket{\psi_{\rm ref}}=\otimes_k\ket{\psi_{\rm ref}}_k$ where $\ket{\psi_{\rm ref}}_k=\alpha\ket{\uparrow}+\beta\ket{\downarrow}$, and the coefficient $\alpha = \cos \left(\frac{\theta}{2}\right)$ and $\beta = e^{i\phi}\sin \left(\frac{\theta}{2}\right)$, we find,
    \begin{align}
  C_{k}(t_{1},t_{2}) = \frac{1}{2} &+ \frac{\sin \theta \cos \phi}{2} \cdot \frac{(t_1-t_2\cos{k})}{|\textbf{d}_{\text{SSH}}(k)|} \notag \\ &+ \frac{\cos \theta}{2}\cdot \frac{t_2\sin{k}}{|\textbf{d}_{\text{SSH}}(k)|}\,.
\end{align}
    Now, following Eq.~\eqref{eq:ck_integral}, the average spread complexity is obtained by integrating over the Brillouin zone, 
\begin{equation}\label{eq:ssh_int_comp}
        C_{\text{SSH}}(t_1,t_2) = \frac{1}{2} + \frac{\sin \theta \cos \phi}{2\pi} \int_0^{\pi}\frac{t_1 - t_2 \cos k}{|\textbf{d}_{\text{SSH}}(k)|}dk\,. 
    \end{equation}
    For ease of notation, we first define,
    \begin{equation}
         \mathcal{I}_{1}(t_{1},t_{2}) = \frac{1}{\pi} \int_{0}^{\pi} dk \frac{t_1-t_2\cos{k}}{\sqrt{t_1^2 + t_2^2 -2t_1t_2\cos{k}}}\,.
    \end{equation}
    Now, the integral $\mathcal{I}_{1}(t_{1},t_{2})$ can be written in terms of the complete elliptic functions of the first and second kind as follows (see Appendix~\ref{appendix:b} for more details),
    \begin{align}
    \mathcal{I}_1(t_1,t_2)=\frac{\delta\,K(m)+s\,E(m)}{\pi t_1}\,,
    \end{align}
    where $\delta = t_{1} - t_{2}$, $s = t_{1} + t_{2}$, and $m = 4t_{1}t_{2}/(t_{1} + t_{2})^{2}$, and the elliptic integrals are given as follows, 
    \begin{align}
K(x) & =\int_{0}^{\pi/2}\frac{dk}{\sqrt{1-x\,\sin^{2}k}}\,, \notag\\
E(x) &=\int_{0}^{\pi/2}dk\sqrt{1-x\,\sin^{2}k}\,.
\end{align}
    We find that for a general reference state $\ket{\psi_{\text{ref}}}$, the spread complexity exhibits a logarithmic divergence at the topological phase boundary ($t_{1} = t_{2}$), i.e., $dC_{\text{SSH}}(t_1,t_2)/dt_2 \sim \log (1/|t_2 - t_1|)$, Fig.~\ref{fig:bloch_spheres_complexity_plots}(c). This follows from the expansion of $\mathcal{I}_1(t_1,t_2)$ near the topological phase transition point, i.e. $(\delta/s)\to 0$,
    \begin{equation}
        \mathcal{I}_{1}(t_{1},t_{2}) = \frac{s}{\pi t_1}+\frac{\delta}{\pi t_1}\ln\frac{4\,s}{|\delta|}
+ \mathcal{O}(\delta^{2}/s^{2}) \,.
    \end{equation}

    However, for $\hat{n}_{\text{ref}}$ chosen such that $\sin \theta \cos \phi = 0$, we get a constant spread complexity and hence no sensitivity to the topological phase transition,  Fig.~\ref{fig:bloch_spheres_complexity_plots}(b). Next, by choosing a momentum-dependent reference state  Bloch vector such that $\hat{n}_{\text{ref}}(k) = (0,\,0,\,+1)$ for $k\geq 0$, and $\hat{n}_{\text{ref}}(k) = (0,\,0,\,-1)$ for $k<0$, we recover the plateau feature of the spread complexity~\cite{PhysRevB.106.195125},  Fig.~\ref{fig:bloch_spheres_complexity_plots}(a). This demonstrates that the behavior of the spread complexity depends on our choice of $\hat{n}_{\text{ref}}(k)$ and it is not universal. Further, for a general reference state, it is the derivative of the spread complexity that shows non-analyticity at the phase boundary.
 
    The above analysis can be generalized to the case where the target state is an excited state. For example, we can choose a target state such that $\hat{n}_{\text{target}}(k) = - \hat{d}_{\text{SSH}}(k)$ for $k\leq 0$ and $\hat{n}_{\text{target}}(k) = \hat{d}_{\text{SSH}}(k)$ for $k>0$, where $\mp \hat{d}_{\text{SSH}}(k)$ are the Bloch vectors corresponding to eigenstates with energy $\mp |\bold{d}_{\text{SSH}}(k)|$. The spread complexity in this case involves incomplete elliptic integrals (see Appendix~\ref{appendix:c}), and is non-analytic at $t_{1} = t_{2}$.   

    \section{Fidelity susceptibility} For a Hamiltonian with a tunable parameter, the fidelity or overlap between the ground states at slightly different values of the parameter varies sharply near a gap closing point. The sensitivity of the ground state to changes in the parameter is quantified by the fidelity susceptibility, which diverges at gap closing points~\cite{doi:10.1142/S0217979210056335, PhysRevA.78.010301, PhysRevA.79.032302}. Since both the Krylov spread complexity and the fidelity susceptibility are defined using overlap between states, this naturally prompts an investigation on the relation between the two quantities.

    For a two-band Hamiltonian $H=\sum_k \bold{d}(k,\lambda)\cdot \boldsymbol{\sigma}$, with a parameter $\lambda$, the fidelity susceptibility is given as follows,
    \begin{equation}\label{eq:fid_susc}
        \chi_F(\lambda) = \frac{1}{2\pi}\int_{-\pi}^{\pi} \chi_F(k,\lambda) = \frac{1}{8\pi}\int_{-\pi}^{\pi} |\partial_{\lambda}\hat{d}(k,\lambda)|^2 dk\,,
    \end{equation}
    where for each momentum mode $k$, the fidelity susceptibility is given by $\chi_F(k,\lambda)=(1/4)\cdot|\partial_{\lambda}\hat{d}(k,\lambda)|^2$ (see Appendix~\ref{appendix:d}). We now consider the derivative of the complexity of the ground state with respect to $\lambda$,
    \begin{equation}\label{eq:der_complexity}
        \partial_{\lambda}C(\lambda) = \sum_{i=1}^{3} \mathcal{Q}_i \int_{-\pi}^{\pi} \partial_{\lambda} \hat{d}_i(k,\lambda) dk\,,
    \end{equation}
    where $i=1,\,2,\,3$ corresponds to $(x,\,y,\,z)$, and $\mathcal{Q}_i$ encodes the information of the reference state in terms of the Bloch sphere angles $\theta$ and $\phi$. We now consider the components $\chi_F^i(\lambda)$ of the fidelity susceptibility, defined as $\chi_F^i(\lambda)=(1/8\pi)\cdot \int_{-\pi}^{\pi} |\partial_{\lambda} \hat{d}_i(k,\lambda)|^2 dk$. By using the Cauchy-Schwartz inequality, we obtain,
\begin{equation}\label{eq:fs_complexity_inequality}
        |\partial_{\lambda}C(\lambda)| \leq 4\pi \sum_{i=1}^3 |\mathcal{Q}_i| \sqrt{\chi_F^i(\lambda)}\,\,.
    \end{equation}
    As a result, any divergence in the derivative of the Krylov spread complexity must be accompanied by a divergence in the fidelity susceptibility. Alternatively, we find that the Krylov spread complexity probes gap closing in the system and not necessarily topological phase transitions.

\begin{figure}[b!]
    \centering
    \includegraphics[width=1.0\linewidth]{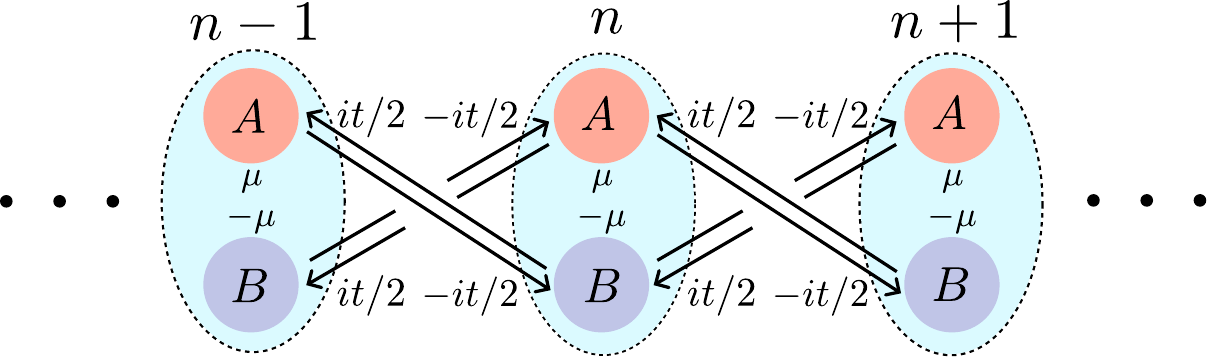}
    \caption{Tight-binding model for the massive Dirac Hamiltonian. Each unit cell contains two sublattices $A$ and $B$, with an imaginary hopping amplitude $it/2$ between the sublattices $A$ and $B$ in adjacent unit cells, and a staggered chemical potential $\mu$ for sublattice $A$, and $-\mu$ for sublattice $B$.}
    \label{fig:massive_dirac}
    \end{figure}

    We now present a model where the gap closing is not accompanied by a topological phase transition. Consider the massive Dirac Hamiltonian,
    \begin{equation}\label{eq:h_md}
        H_{\text{MD}}(\mu) = \sum_k \left ( t \sin k \,\sigma_x + \mu \,\sigma_z \right )\,,
    \end{equation}
    where $t$ is the hopping amplitude, which we set to 1, and $\mu$ is the staggered chemical potential, see Fig.~\ref{fig:massive_dirac}. The Zak phase~\cite{6ee78b01-9610-3faf-a626-578be4e5d3b4,Asboth2016} is identically zero across the gap closing point at $\mu=0$, and hence it is not a topological phase transition. We point out that the Hamiltonian for the Cooper pair box~\cite{Wong2025QuantumComputingArchitecture} takes the form of Eq.~(\ref{eq:h_md}). In this case the magnetic flux $\Phi$ plays the role of the momentum and the charging energy mimics the staggered chemical potential (see Appendix~\ref{appendix:d}). This demonstrates that our analysis is not restricted to lattice models and is applicable to any system with Hamiltonian of the form $H(k,\,\lambda) = \bold{d}(k,\,\lambda)\cdot \boldsymbol{\sigma}$, where $k$ is any periodic variable.
    
    For calculating the Krylov spread complexity for the ground state of the massive Dirac Hamiltonian, we start with the choice $\hat{n}_{\text{ref}}=(\sin \theta \cos \phi,\, \sin \theta \sin \phi,\, \cos \theta)$. By using Eq.~\eqref{eq:ck_overlap_bloch}, we find, 
    \begin{equation}\label{eq:md_ham_complexity}
        C_{\text{MD}}(\mu) = \frac{1}{2} + \frac{\mu \cos \theta}{\pi\sqrt{1+\mu^2}}K\left ( \frac{1}{1+\mu^2} \right )\,,
    \end{equation}
    where $K(x)$ is the complete elliptic integral of the first kind~\cite{gradshteyn2007table}. Near the gap closing point $\mu=0$, we find that the derivative of the complexity has a logarithmic divergence, i.e., $\partial_{\mu} C_{\text{MD}}(\mu) \sim \log (1/|\mu|)$ as $\mu \to 0$. The derivative of the complexity diverges at a gap closing point, that does not correspond to a topological phase transition. Further, the fidelity susceptibility is obtained using Eq.~(\ref{eq:fid_susc}) and is given as (see Appendix~\ref{appendix:d}),
    \begin{equation}\label{eq:md_ham_fs}
        \chi_F^{\text{MD}}(\mu) = \frac{1}{8|\mu|(1+\mu^2)^{3/2}}\,,
    \end{equation}
    which diverges near the gap closing point, i.e., $\chi_F^{\text{MD}}(\mu) \sim 1/|\mu|$ as $\mu \to 0$. The fidelity susceptibility diverges faster than the derivative of complexity. Similarly, for the SSH model, fidelity susceptibility diverges at the gap closing point ($t_{1} = t_{2}$) that is $\chi_F^{\text{SSH}}(t_1,t_2) \sim 1/|t_1 - t_2|$, whereas from Eq.~(\ref{eq:ssh_int_comp}), we have $\partial_{t_2}C_{\text{SSH}}(t_1,t_2) \sim \log (1/|t_1 - t_2|)$. Note that the only non-vanishing term in $\partial_{t_{2}} C_{\text{SSH}}(t_{1},t_{2})$, Eq.~(\ref{eq:der_complexity}) for the SSH model is $\partial_{t_{2}} C_{\text{SSH}}^{(1)}(t_{1},t_{2}) = \mathcal{Q}_{1}\int_{-\pi}^{\pi}\partial_{t_{2}} \hat{d}_{\text{SSH}}^{(1)}(k) dk$, whereas for the massive Dirac Hamiltonian it is $\partial_{\mu} C_{\text{MD}}^{(3)}(\mu) = \mathcal{Q}_{3}\int_{-\pi}^{\pi}\partial_{\mu} \hat{d}_{\text{MD}}^{(3)}(k) dk$. In both cases, the ratio $R(\lambda) = |\partial_{\lambda}C^{(i)}(\lambda)|/4\pi|\mathcal{Q}_i|\sqrt{\chi_F^i(\lambda)}$ saturates to a constant value $\sqrt{2/3}$ in the limit $\lambda \to \infty$ (see Appendix~\ref{appendix:d}).

    \section{Duality in the SSH model} 
\begin{figure}[htbp]
    \centering
    \includegraphics[width=1.0\linewidth]{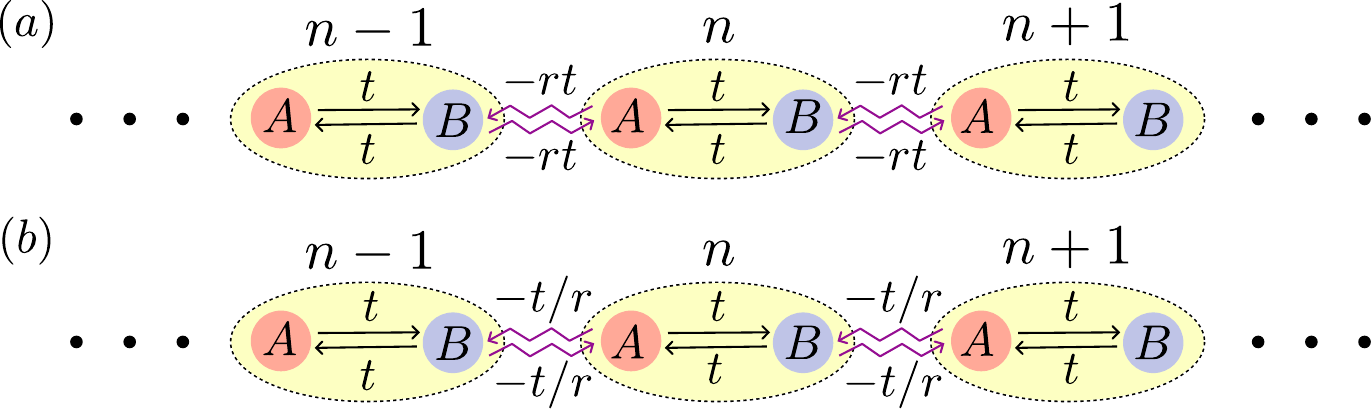}
    \caption{Dual SSH models with intracell coupling $t$, and (a) intercell coupling $-rt$, (b) intercell coupling $-t/r$.}
    \label{fig:ssh_duality}
    \end{figure}

The study of Krylov spread complexity and fidelity susceptibility reveals a duality in the SSH model. In the SSH model, Eq.~(\ref{eq:ssh_hamiltonian}), the ratio $t_2/t_1$ determines the topological phase of the system. We now parametrize the SSH model using $t$, which will be fixed to be the intracell coupling, and a dimensionless scalar $r$. The duality is between two class of SSH Hamiltonians related via $r \leftrightarrow 1/r$, as shown in Fig.~\ref{fig:ssh_duality}. These two Hamiltonians are given by $H_{\text{I}}(k) = (t-rt)\cos k \, \sigma_x + rt \sin k \, \sigma_y$ and $H_{\text{II}}(k) = \left[ t - (t/r) \right]  \cos k \, \sigma_x + (t/r) \sin k \, \sigma_y$. They are related via a non-unitary transformation,
    \begin{equation}\label{eq:non_inv_mapping}
        H_{\text{II}}(k) = \frac{1}{r} U(k) H_{\text{I}}(k) U^{\dagger}(k)\,, 
    \end{equation}
    where $U(k)  = \text{diag}\{(r-e^{ik})/(1-re^{ik}),\,1\}$ is a unitary matrix. At the self-dual point $r=1$, $H_{\text{I}}(k)$ and $H_{\text{II}}(k)$ become identical and $U(k)$ is the identity matrix. This duality divides the entire parameter space of the SSH model into two regions (trivial and topological) and draws a mapping between them.

    Calculating the winding number~\cite{6ee78b01-9610-3faf-a626-578be4e5d3b4,Asboth2016}, we find that for $H_{\text{I}}$, the winding number $\nu_{\text{I}}$ is 0 for $r<1$ and 1 for $r>1$. Correspondingly, for $H_{\text{II}}$, the winding number $\nu_{\text{II}}$ is 1 for $r<1$ and 0 for $r>1$. It is more instructive to study the fidelity susceptibility in these two cases. Denoting the fidelity susceptibility for $H_{\text{I}}$ as $\chi_F^{(\text{I})}(r)$ and for $H_{\text{II}}$ as $\chi_F^{(\text{II})}(1/r)$, using Eq.~(\ref{eq:fid_susc}), we find (see Appendix~\ref{appendix:e}),
    \begin{equation}\label{eq:fs_duality}
        \chi_F^{(\text{II})}(1/r) = r^4 \chi_F^{(\text{I})}(r).
    \end{equation}
    A similar relation also holds for the complexity of the ground states of $H_{\text{I}}$ and $H_{\text{II}}$, denoted by $C^{(\text{I})}$ and $C^{(\text{II})}$ (see Appendix~\ref{appendix:e}),
    \begin{equation}\label{eq:complexity_duality}
        C^{(\text{II})}(1/r) = \frac{C^{(\text{I})}(r) - \mathcal{H}(r)}{r},
    \end{equation}
    where $\mathcal{H}(r) = \frac{(1-r)}{2\pi}\left [ \pi + 2 \sin \theta \cos \phi K \left ( 4r/(1+r)^2 \right ) \right ]$, $\theta,\, \phi$ are the reference state Bloch sphere angles, and $K(x)$ is the complete elliptic integral of the first kind. Hence, from Eq.~(\ref{eq:fs_duality}) and Eq.~(\ref{eq:complexity_duality}), the knowledge of fidelity susceptibility and the ground state spread complexity in one phase of the SSH model is sufficient to determine these in the other phase. The ratio $R(r)$ for the $\text{SSH}$ model is also invariant under $r \leftrightarrow 1/r$, which is a clear manifestation of the duality (see Appendix~\ref{appendix:e}).

    \section{Discussion} Our analysis establishes that the derivative of the Krylov spread complexity for the ground state is bounded by the fidelity susceptibility, Eq.~\eqref{eq:fs_complexity_inequality} in two-band Hamiltonians. The spread complexity is obtained using a purely geometric formulation based on overlaps of Bloch sphere vectors, Eq.~\eqref{eq:ck_overlap_bloch} without explicitly constructing the circuit Hamiltonian. Any divergence in the derivative of the spread complexity also implies divergence in the fidelity susceptibility. Therefore, the derivative of the complexity is a probe of gap closings, and not just topological phase transitions. As an example, we work out spread complexity in the massive Dirac Hamiltonian exhibiting non-analyticity at the gap closing point, Eq.~\eqref{eq:md_ham_complexity} which is not a topological phase transition point. Further, in the SSH model, we discuss a non-unitary duality, Eq.~\eqref{eq:non_inv_mapping} relating the two phases, which manifests itself in the fidelity susceptibility, Eq.~\eqref{eq:fs_duality} and the ground state Krylov spread complexity, Eq.~\eqref{eq:complexity_duality}.

    We extend this work to the non-Hermitian SSH model~\cite{PhysRevLett.121.086803} with periodic boundary conditions, and show that spread complexity is also sensitive to the gap closing point in this setting (see Appendix~\ref{appendix:f}). Extending this machinery to the non-Hermitian SSH model under open boundary conditions will be an important step towards integrating the notion of complexity with non-Hermitian topological phenomena such as the skin effect. However, this is a challenging problem because now the Krylov subspace is not $2-$dimensional, and we cannot exploit Bloch sphere geometry anymore.

    It would also be interesting to explore how the first derivative of Krylov spread complexity is related to the fidelity susceptibility for a general system, other than two-band Hamiltonians. Establishing such a connection would provide further insight into the geometric  content encoded by Krylov spread complexity~\cite{Caputa:2021sib,PhysRevResearch.6.L042001,Cafaro2026KrylovsSC}. Setting a lower bound on the derivative of the complexity is an intriguing problem, and extending our analysis to two-band Hamiltonians in two or higher spatial dimensions, such as Chern insulators and graphene~\cite{6ee78b01-9610-3faf-a626-578be4e5d3b4,Asboth2016}, may provide some insights on such a lower bound.

\section*{A\MakeLowercase{cknowledgments}} It is a pleasure to thank Claudio Chamon for insightful questions, comments and discussions. We would also like to thank Jason Alicea and Julian May-Mann for useful discussions on topological phases of matter. R.C. gratefully acknowledges the Summer Research Grant awarded by Department of Physics and Astronomy, Purdue University. S.R. and S.S.S. are partially supported by Department of Physics and Astronomy, Purdue University through Graduate Assistantship.

\begin{appendix}
\begin{widetext}

\setcounter{equation}{0}
\setcounter{figure}{0}
\setcounter{table}{0}
\setcounter{section}{0}

\makeatletter
\renewcommand{\thesection}{\Alph{section}}
\renewcommand{\thesubsection}{\arabic{subsection}}

\renewcommand{\theequation}{\Alph{section}\arabic{equation}}
\renewcommand{\thetable}{\Alph{section}\arabic{table}}
\renewcommand{\thefigure}{\Alph{section}\arabic{figure}}

\@addtoreset{equation}{section}
\@addtoreset{table}{section}
\@addtoreset{figure}{section}
\makeatother

\begin{center}
\textbf{
Supplemental Material: Krylov complexity and fidelity susceptibility in two-band Hamiltonians
}
\end{center}

In this supplementary material, we show the details of the results discussed in the article. We start with Appendix~\ref{appendix:a}, where we have the details of the Su-Schrieffer-Heeger (SSH) model and the massive Dirac Hamiltonian. By calculating the Zak phase (winding number), we show the topological nature of SSH model and that the gap closing in massive Dirac Hamiltonian is not accompanied by topological phase transition. In Appendix~\ref{appendix:b}, we give the details of obtaining the Krylov spread complexity in two-band Hamiltonians using Bloch sphere data. We derive the spread complexity for SSH model. Important formulas for elliptic functions are also summarized. Further we derive the asymptotic behavior of the spread complexity. In Appendix~\ref{appendix:c}, we have derived the spread complexity for a general (excited) eigenstate of the SSH model. The following Appendix~\ref{appendix:d} includes all the details for deriving fidelity susceptibility for two-band Hamiltonians. We give full derivation for obtaining the inequality between the Krylov spread complexity and the fidelity susceptibility. Further we also show that spread complexity in the massive Dirac Hamiltonian is non-analytic at the gap closing point, and is not accompanied by topological phase transition. We explicitly show the inequality also for SSH model. Finally, we establish equivalence at the Hamiltonian level between massive Dirac model and Cooper pair box. In Appendix~\ref{appendix:e}, we give the details of the non-unitary duality between topological and trivial phase in the the SSH model. In the last Appendix~\ref{appendix:f}, we derive the spread complexity for non-Hermitian SSH model using the Bloch sphere data, and show the non-analyticity at the gap closing point.


\section{Su-Schrieffer-Heeger (SSH) model with nearest neighbor hoppings}  \label{appendix:a}

The Su-Schrieffer-Heeger (SSH) model is a tight binding Hamiltonian where we have two sites in a given unit cell with inter-cell and intra-cell hoppings\cite{Asboth2016}. The Hamiltonian in the real space with $L$ number of sites is given as follows,
\begin{equation}
H_{\text{SSH}}=\sum_{n=1}^{L}\left( t_{1}c_{A,n}^{\dagger}c_{B,n}+t_{1}c_{B,n}^{\dagger}c_{A,n}-t_{2}c_{B,n}^{\dagger}c_{A,n+1}-t_{2}c_{A,n+1}^{\dagger}c_{B,n} \right) \,,
\end{equation}
where $t_{1}$ is the inter-cell hopping amplitude and $t_{2}$ is the intra-cell hopping amplitude. Next, we Fourier transform the SSH Hamiltonian,
\begin{align}\label{eq:ft_creation_annihi_ops}
c_{n,A} & =\frac{1}{\sqrt{L}}\sum_{k}e^{-ikn}d_{k,A}\,, \text{ } c_{n,B} =\frac{1}{\sqrt{L}}\sum_{k}e^{-ikn}d_{k,B}\,.
\end{align}
This leads to the following,
\begin{equation}
\begin{split}
& \sum_{n}t_{1}c_{A,n}^{\dagger}c_{B,n} =\frac{1}{L}\sum_{k_{1}k_{2}}\sum_{n}e^{ik_{1}n}e^{-ik_{2}n}d_{k_{1},A}^{\dagger}d_{k_{2},B}=\sum_{k}t_{1}d_{k,A}^{\dagger}d_{k,B}, \\
& \sum_{n}t_{2}c_{B,n}^{\dagger}c_{A,n+1} =\frac{1}{L}\sum_{k_{1}k_{2}}\sum_{n}e^{ik_{1}n}e^{-ik_{2}n-ik_{2}}d_{k_{1},B}^{\dagger}d_{k_{2},A}=\sum_{k}t_{2}e^{-ik}d_{k,B}^{\dagger}d_{k,A}.
\end{split}
\end{equation}
This implies, we are going to have the following,
\begin{equation} \label{eq: SSH Hamiltonian k-space summed over}
\begin{split}
H_{\text{SSH}} & =\sum_{k}t_{1}d_{k,A}^{\dagger}d_{k,B}+\sum_{k}t_{1}d_{k,B}^{\dagger}d_{k,A}-\sum_{k}t_{2}e^{-ik}d_{k,B}^{\dagger}d_{k,A}-\sum_{k}t_{2}e^{ik}d_{k,A}^{\dagger}d_{k,B}\\
 & =\sum_{k}\left(\begin{array}{cc}
d_{k,A}^{\dagger} & d_{k,B}^{\dagger}\end{array}\right)\left[\begin{array}{cc}
0 & t_{1}-t_{2}e^{ik}, \\
t_{1}-t_{2}e^{-ik} & 0
\end{array}\right]\left(\begin{array}{c}
d_{k,A}\\
d_{k,B}
\end{array}\right).
\end{split}
\end{equation}
Next, let us write the Hamiltonian in term of the Pauli matrices,
\begin{equation}
\begin{split}
H_{\text{SSH}}(k) & =\left[\begin{array}{cc}
0 & t_{1}-t_{2}e^{ik}\\
t_{1}-t_{2}e^{-ik} & 0
\end{array}\right] = \left[\begin{array}{cc}
0 & t_{1}-t_{2}\cos\,k - i\,t_{2}\,\sin\,k \\
t_{1}-t_{2}\cos\,k + i\,t_{2}\,\sin\,k  & 0
\end{array}\right] \, \\
& = \left[\begin{array}{cc}
0 & t_{1}-t_{2}\cos\,k\\
t_{1}-t_{2}\cos\,k & 0
\end{array}\right] + \left[\begin{array}{cc}
0 & -i\,t_{2}\sin\,k\\
i\,t_{2}\sin\,k & 0
\end{array}\right]\, \\
& = (t_{1} - t_{2}\cos\,k)\,\sigma_{x} + t_{2}\sin\,k\,\sigma_{y}\,.
\end{split}
\end{equation}
Hence, we have the following for the SSH model,
\begin{align}\label{eq:ssh_ham_appen_a}
    H_{\text{SSH}}(k) =  \tilde{d}_{x}(k)\,\sigma_{x} + \tilde{d}_{y}(k)\,\sigma_{y}\,,
\end{align}
where,
\begin{align}
    \tilde{d}_{x}(k)= t_{1} - t_{2}\cos\,k\,, \quad \text{and} \quad \tilde{d}_{y}(k) =  t_{2}\sin\,k \,.
\end{align}
At this point for convenience, we introduce a unitary transformation of the basis $(\sigma_x, \sigma_y, \sigma_z) \to (\sigma_x, \sigma_z, -\sigma_y)$. This implies that the Hamiltonian in the new basis is given as follows,
\begin{align}\label{eq:ssh_ham_appen_a}
    h(k) = \left[\begin{array}{cc}
t_2 \sin k & t_{1}-t_{2}\cos\,k\\
t_{1}-t_{2}\cos\,k & - t_2 \sin k
\end{array}\right] = d_{x}(k)\sigma_{x} + d_{z}(k) \sigma_{z} \,,
\end{align}
where we have,
\begin{align}
    d_{x}(k)= t_{1} - t_{2}\cos\,k\,, \quad \text{and} \quad d_{z}(k) =  t_{2}\sin\,k \,.
\end{align}
After diagonalization, we obtain the eigenstates of the SSH Hamiltonian, see Eq.~\eqref{eq:ssh_ham_appen_a},
\begin{align}\label{eq:GS vector in k-space}
    \ket{-\textbf{E}_k} &= \frac{1}{\sqrt{2 E_k (E_k + d_z(k))}} \begin{pmatrix}
        -d_x(k) \\
        E_k + d_z(k)
    \end{pmatrix}\,, \, \, \text{with eigenvalue } -E_{k} = -\sqrt{t_1^2 + t_2^2 - 2 t_1 t_2 \cos k}\,, \\
    \ket{\textbf{E}_k} &= \frac{1}{\sqrt{2 E_k (E_k + d_z(k))}} \begin{pmatrix}
        E_k + d_z(k) \\
        d_x(k)
    \end{pmatrix}\,, \, \, \text{with eigenvalue } E_{k} = \sqrt{t_1^2 + t_2^2 - 2 t_1 t_2 \cos k}\,.
\end{align}
From here we can write down the corresponding Bloch sphere vectors,
\begin{align}
    -\hat{d}_{\text{SSH}}(k) &= -\frac{1}{\sqrt{t_{1}^{2} + t_{2}^{2} - 2t_{1}t_{2}\cos\,k}}(t_{1} - t_{2}\cos\,k\,, 0\,,t_{2}\sin\,k )\,, \, \text{for the eigenstate } \ket{-\textbf{E}_k}\,,\\
    \hat{d}_{\text{SSH}}(k) &= \frac{1}{\sqrt{t_{1}^{2} + t_{2}^{2} - 2t_{1}t_{2}\cos\,k}}(t_{1} - t_{2}\cos\,k\,, 0\,,t_{2}\sin\,k )\,, \, \text{for the eigenstate } \ket{\textbf{E}_k}\,,
\end{align}

\subsection{Winding number in Su-Schrieffer-Heeger (SSH) model}\label{subsec:winding_ssh}
The topological property of the Su-Schriefer-Heeger (SSH) model is captured by defining a winding number or equivalently topological invariant~\cite{6ee78b01-9610-3faf-a626-578be4e5d3b4}. This is a manifestation of the Bloch bulk-boundary correspondence, the winding number is calculated using an infinite lattice (hence a bulk theory) which determines whether the system hosts edge states (living on the boundary) in a finite system. Let us start with the Hamiltonian of the SSH model in the momentum space,
\begin{equation}
H_{\text{SSH}}=\left[\begin{array}{cc}
0 & t_{1}-t_{2}e^{ik}\\
t_{1}-t_{2}e^{-ik} & 0
\end{array}\right]=\left[\begin{array}{cc}
0 & f_{1}(k)\\
f_{1}(k)^{*} & 0
\end{array}\right]\,.
\end{equation}
Winding number (or the Zak phase) is given as follows,
\begin{align}
\nu & =\frac{1}{2\pi i}\int_{-\pi}^{\pi}dk\partial_{k}\ln f(k) = \frac{1}{2\pi i}\int_{-\pi}^{\pi}dk\frac{-it_{2}e^{ik}}{t_{1}-t_{2}e^{ik}} =\frac{1}{2\pi}\int_{-\pi}^{\pi}dk\frac{-t_{2}e^{ik}}{t_{1}-t_{2}e^{ik}}\,,
\end{align}
Now let us make the substitution as follows,
\begin{equation}
z=e^{ik}\Rightarrow dz=ie^{ik}dk \,.
\end{equation}
This implies,
\begin{equation}
\nu=\frac{1}{2\pi}\oint_{|z|=1}\frac{dz}{iz}\frac{-t_{2}z}{t_{1}-t_{2}z}=\frac{1}{2\pi i}\oint_{|z|=1}dz\frac{1}{z-\frac{t_{1}}{t_{2}}} = \begin{cases}
    0\,,\qquad(\text{for }t_{1}/t_{2}>1) \\
    1\,,\qquad(\text{for }t_{1}/t_{2}<1)
\end{cases} \,,
\end{equation}
and hence we find that with vanishing winding number, we do not have edge states in the system but winding number $1$ corresponds to topological phase with edge states.

\subsection{Massive Dirac Hamiltonian }
In this subsection, we are going to study one dimensional massive Dirac (MD) Hamiltonian where we see a gap closing in the spectrum but this does not correspond to any topological phase transition. The MD Hamiltonian is given as follows,
\begin{equation}
\begin{split}
    H_{\text{MD}} &= \sum_{n=1}^{L} \Big{(} \mu \,c^{\dagger}_{A,n}c_{A,n} - \mu\, c^{\dagger}_{B,n}c_{B,n}  \, \\
     & \quad +  \frac{t}{2}\,e^{i\pi/2}\,c^{\dagger}_{A,n}c_{B,n+1} + \frac{t}{2}\,e^{-i\pi/2}\,c_{B,n+1}^{\dagger}c_{A,n}+  \frac{t}{2}\,e^{i\pi/2}\,c^{\dagger}_{B,n}c_{A,n+1} + \frac{t}{2}\,e^{-i\pi/2}\,c_{A,n+1}^{\dagger}c_{B,n} \Big{)}.
\end{split}
\end{equation}
By using the Fourier transform definition we used previously for SSH Hamiltonian, Eq.~\eqref{eq:ft_creation_annihi_ops} we find the following,
\begin{align}
\sum_{n}\frac{t}{2}\,c_{A,n}^{\dagger}c_{B,n+1} & =\frac{1}{L}\sum_{k_{1}k_{2}}\sum_{n}e^{ik_{1}n}e^{-ik_{2}n}e^{-ik_{2}}\,\frac{t}{2}\,d_{k_{1},A}^{\dagger}d_{k_{2},B}=\sum_{k}e^{-ik}\,\frac{t}{2}\,d_{k,A}^{\dagger}d_{k,B}\,,
\end{align}
and from here we have the following for the Hamiltonian in the momentum space,
\begin{equation}
\begin{split}
    H_{\text{MD}} &= \sum_{k} \Big{(} \mu\,d^{\dagger}_{A,k}d_{A,k} - \mu\,d^{\dagger}_{B,k}d_{B,k} + \frac{t}{2}\,e^{i\left(\frac{\pi}{2} - k\right)}\, d_{A,k}^{\dagger}d_{B,k} + \frac{t}{2}\,e^{-i\left(\frac{\pi}{2} - k\right)}\, d_{B,k}^{\dagger}d_{A,k} \, \\
    & \qquad + \frac{t}{2}\,e^{i\left(\frac{\pi}{2} - k\right)}\, d_{B,k}^{\dagger}d_{A,k} + \frac{t}{2}\,e^{-i\left(\frac{\pi}{2} - k\right)}\, d_{A,k}^{\dagger}d_{B,k} \Big{)}\, \\
    &= \sum_{k} \left( \mu\,d^{\dagger}_{A,k}d_{A,k} - \mu\,d^{\dagger}_{B,k}d_{B,k} + t\,\sin k \, d_{A,k}^{\dagger}d_{B,k} + t\,\sin k\, d_{B,k}^{\dagger}d_{A,k} \right)\,.
\end{split}
\end{equation}
We can write the above Hamiltonian more succinctly as follows,
\begin{align}
    H_{\text{MD}} & =\sum_{k}\left(\begin{array}{cc}
d_{k,A}^{\dagger} & d_{k,B}^{\dagger}\end{array}\right)\left[\begin{array}{cc}
\mu & t\,\sin k\\
t\,\sin k & -\mu
\end{array}\right]\left(\begin{array}{c}
d_{k,A}\\
d_{k,B}
\end{array}\right)\,.
\end{align}
In terms of the Pauli matrices, this Hamiltonian takes the form
\begin{align}\label{eq:md_ham_k}
    H_{\text{MD}}(k) = t\,\sin k \sigma_{x} + \mu \,\sigma_{z}\,.
\end{align}
Thus, we see that the gap closes at $k=0,\, \pm \pi$ and $\mu = 0$.

\subsection{Winding number in the massive Dirac Hamiltonian }\label{subsubsec:massive_dirac_winding}
In this subsection, we are going to show that the massive Dirac Hamiltonian, Eq.~\eqref{eq:md_ham_k} does not admit topological phase transition as we tune the mass term ($\mu$ in Eq.~\eqref{eq:md_ham_k}) across the gap closing point. In order to do this, we explicitly evaluate the winding number defined as follows,
\begin{align}\label{eq:winding_number_raw_def}
    \nu = \frac{1}{2\pi}\,\int_{-\pi}^{\pi} \left( \hat{d}(k)\, \times\, \frac{d}{dk}\hat{d}(k) \right)_{z} dk\,,
\end{align}
where we have defined,
\begin{align}\label{eq:md_d}
    \hat{d}(k) = \left( \frac{t\,\sin k}{\sqrt{t^{2}\,\sin^{2}k\,+\,\mu^{2}}},\, 0,\, \frac{\mu}{\sqrt{t^{2}\,\sin^{2}k\,+\,\mu^{2}}} \right)\,.
\end{align}
From here, we find,
\begin{align}\label{eq:md_d_derivative}
    \frac{d}{dk}\hat{d}(k) = \left( \frac{t\,\cos k}{\sqrt{t^{2}\,\sin^{2}k\,+\,\mu^{2}}}-\frac{t^{3}\,\sin^{2}k\,\cos k}{(t^{2}\,\sin^{2}k\,+\,\mu^{2})^{3/2}},\, 0,\, -\frac{\mu\,t^{2}\,\sin k\,\cos k}{\sqrt{t^{2}\,\sin^{2}k\,+\,\mu^{2}}} \right)\,.
\end{align}
Next we note from Eq.~\eqref{eq:winding_number_raw_def} that for obtaining the winding number we just need to focus on the $z-$component of the cross product, and hence, we find using Eq.~\eqref{eq:md_d} and Eq.~\eqref{eq:md_d_derivative},
\begin{align}
    \left( \hat{d}(k)\, \times\, \frac{d}{dk}\hat{d}(k) \right)_{z} = 0 \,.
\end{align}
Hence, the system is trivial no matter what value does the mass gap ($\mu$) takes or tuned across the gap closing point. It is a trivial system where we have gap closing but topological phase transition.

\section{Krylov spread complexity in two-band Hamiltonians }\label{appendix:b}
In this subsection, we discuss the Krylov spread complexity for a two-level Hamiltonian of the form $H(k)= \mathbf{d}(k) \cdot \boldsymbol{\sigma}$. Note that in quantum systems with $2-$dimensional Hilbert space, the Krylov subspace is also $2-$dimensional. The Krylov subspace is generated by,
\begin{equation}
\mathbb{K}^{2} = \mathrm{span} \{ |\mathcal{K}_0\rangle, \, \mathcal{H} |\mathcal{K}_0\rangle \}\,,
\end{equation}
where $|\mathcal{K}_0\rangle$ is a reference state which is an arbitrary quantum state in the $2-$dimensional Hilbert space and the operator $\mathcal{H}$ is the circuit Hamiltonian that takes the reference state to a target state $|\psi_{\text{target}}\rangle$ in unit time. In general, the reference state $|\mathcal{K}_{0}\rangle$ is not an eigenstate of the circuit Hamiltonian $\mathcal{H}$. In order to obtain the Krylov basis, one needs to perform a Gram-Schimdt orthogonalization on the vectors $\{ |\mathcal{K}_0\rangle, \, \mathcal{H} |\mathcal{K}_0\rangle \}$, giving us,
\begin{align}
    \mathbb{K}^{2} = \mathrm{span} \{ |\mathcal{K}_0\rangle, \, |\mathcal{K}_1\rangle \} ; \text{ } \langle \mathcal{K}_0 | \mathcal{K}_1\rangle = 0 \,.
\end{align}
In a 2-dimensional Hilbert space, the state orthogonal to any given state can be uniquely determined up to an overall phase. Hence, one does not explicitly require the circuit Hamiltonian to construct the Krylov basis. Thus, once we choose a reference state $|\mathcal{K}_{0}\rangle$, we can automatically determine $|\mathcal{K}_{1}\rangle$ up to an overall phase.  Let us choose our reference state to be a generic state on the Bloch sphere given by,
\begin{equation}
|\mathcal{K}_0\rangle = \alpha |\uparrow\rangle + \beta |\downarrow\rangle\,,
\end{equation}
where $|\alpha|^2 + |\beta|^2 = 1$, and $|\uparrow\rangle$ and $|\downarrow\rangle$ are eigenstates of $\sigma_{z}$ with eigenvalues $+1$, and $-1$ respectively,
\begin{equation}\label{eq:arrow_kets_appen_b}
|\uparrow\rangle \equiv \begin{pmatrix} 1 \\ 0 \end{pmatrix}
\text{, }
|\downarrow\rangle \equiv \begin{pmatrix} 0 \\ 1 \end{pmatrix}\,.
\end{equation}
Since the space is two-dimensional, we make the following choice for the other element of the Krylov basis vector, $|\mathcal{K}_1\rangle$,
\begin{equation}
|\mathcal{K}_{1}\rangle
= -\beta^* |\uparrow\rangle + \alpha^* |\downarrow\rangle,
\end{equation}
and we have,
\begin{equation}
\langle \mathcal{K}_0 | \mathcal{K}_1 \rangle = 0\,.
\end{equation}
Next, we can start calculating the Krylov spread complexity for a given target state $|\Omega\rangle$. First step is to expand the target state in the Krylov basis as follows,
\begin{equation}
    |\Omega\rangle = \psi_{0}|\mathcal{K}_{0}\rangle + \psi_{1}|\mathcal{K}_{1}\rangle\,, \quad \text{where} \quad \psi_{0} = \langle\mathcal{K}_{0}|\Omega\rangle \quad \text{and} \quad \psi_{1} = \langle\mathcal{K}_{1}|\Omega\rangle\,.
\end{equation}
From here, we directly obtain the Krylov spread complexity as follows,
\begin{align}
    \mathcal{C} = \sum_{n=0}^{1} n|\psi_{n}|^{2} = |\psi_{1}|^{2} = 1 - |\psi_{0}|^{2} = 1 - |\langle\mathcal{K}_{0}|\Omega\rangle|^{2}\,,
\end{align}
Additionally, we can use the following relation between inner product of quantum states and dot product of the corresponding Bloch vectors,  
\begin{align}
    |\langle \psi_{1}  | \psi_{2}\rangle|^{2} = \frac{1+\hat{n}_{\psi_{1}}\cdot \hat{n}_{\psi_{2}}}{2}\,,
\end{align}
where we have,
\begin{align}
    |\psi_{1}\rangle = \cos \frac{\theta_{1}}{2}|\uparrow\rangle + e^{i\phi_{1}}\sin\frac{\theta_{1}}{2}|\downarrow\rangle \qquad \text{and} \qquad |\psi_{1}\rangle = \cos \frac{\theta_{2}}{2}|\uparrow\rangle + e^{i\phi_{2}}\sin\frac{\theta_{2}}{2}|\downarrow\rangle\,,
\end{align}
and the corresponding unit vectors on the Bloch sphere are given as follows,
\begin{align}
    \hat{n}_{\psi_{1}} = (\cos \phi_{1}\sin \theta_{1},\sin \phi_{1}\sin \theta_{1},\cos \theta_{1}) \qquad \text{and} \qquad \hat{n}_{\psi_{2}} = (\cos \phi_{2}\sin \theta_{2},\sin \phi_{2}\sin \theta_{2},\cos \theta_{2})\,,
\end{align}
Hence, the complexity per momentum mode $\textit{k}$ can simply be written as follows,
\begin{align}\label{eq:C_k}
    C_k = \frac{1-\hat{n}_{\mathcal{K}_{0}}\cdot \hat{n}_{\Omega}(k)}{2} = \frac{1-\hat{n}_{\text{ref}}\cdot \hat{n}_{\text{target}}(k)}{2}\,.
\end{align}
We clearly see an advantage in exploiting the fact that the Hilbert space is two-dimensional, and Bloch sphere properties to obtain the Krylov spread complexity without introducing any coherent state formalism. From here, we can obtain the averaged Krylov spread complexity density,
\begin{align}\label{eq:Integrated Complexity definition}
    C = \frac{1}{2\pi} \int_{-\pi}^{\pi} dk\, C_{k}\,.
\end{align}

\subsection{Spread complexity for a general reference state for SSH }

In this section we are going to evaluate the Krylov spread complexity for the ground state of SSH model Eq.~\eqref{eq:groundstate_ssh_appen_b}, by exploiting the fact that Hilbert space of SSH model is two dimensional. For the SSH model, ground state as a function of momentum $k\in [-\pi,\pi]$ is given as follows, which is also our target state,
\begin{align}\label{eq:groundstate_ssh_appen_b}
    \hat{n}_{\text{target}}(k) = -\left(
\frac{t_1-t_2\cos{k}}{\sqrt{t_1^2 + t_2^2 -2t_1t_2\cos{k}}},
\,0,\,
\frac{t_2\sin{k}}{\sqrt{t_1^2 + t_2^2 -2t_1t_2\cos{k}}}
\right)\,,
\end{align}
and let us choose the \textit{k-independent} reference state to be the following,
\begin{align}\label{eq:generic_ref_state}
    \hat{n}_{\text{ref}} = (\cos \phi\sin \theta,\,\sin \phi\sin \theta,\,\cos \theta)\,.
\end{align}
Note that the above reference state in Eq.~\eqref{eq:generic_ref_state} is equivalent to the ket $|\psi_{\text{ref}}\rangle$ such that,
\begin{align}\label{eq:ket_ref}
    |\psi_{\text{ref}}\rangle \equiv \begin{pmatrix} \alpha \\ \beta \end{pmatrix}\,, \qquad \alpha = \cos \frac{\theta}{2}\,, \quad \beta= e^{i\phi} \sin \frac{\theta}{2}\,.
\end{align}
Next, note that for each momentum mode $k\in[-\pi,\pi]$ we are going to have different Krylov spread complexity, or in other words Krylov spread complexity is going to be a function of momentum in addition to hopping amplitudes $t_{1}$ and $t_{2}$ (in SSH Hamiltonian) and is given as follows,
\begin{equation}
C_{k}(t_{1},t_{2}) =\frac{1-\hat{n}_{\text{ref}}\cdot \hat{n}_{\text{target}}(k)}{2} \,.
\end{equation}
This implies Krylov spread complexity for each momentum mode $k$ is given as follows,
\begin{align}
  C_{k}(t_{1},t_{2}) = \frac{1}{2} + \frac{\sin \theta \cos \phi}{2} \frac{t_1-t_2\cos{k}}{\sqrt{t_1^2 + t_2^2 -2t_1t_2\cos{k}}} + \frac{\cos \theta}{2}\frac{t_2\sin{k}}{\sqrt{t_1^2 + t_2^2 -2t_1t_2\cos{k}}}\,.
\end{align}
Next, we define an momentum averaged Krylov spread complexity over the entire Brillouin zone i.e.,
\begin{equation}
\begin{split}
    C(t_{1},t_{2}) = \frac{1}{2\pi} \int_{-\pi}^{\pi} dk \, C_{k}(t_{1},t_{2}) &= \frac{1}{2} + \frac{\sin \theta \cos \phi}{4\pi} \int_{-\pi}^{\pi} dk\,\frac{t_1-t_2\cos{k}}{\sqrt{t_1^2 + t_2^2 -2t_1t_2\cos{k}}}+ \frac{\cos \theta}{4\pi}\int_{-\pi}^{\pi} dk \, \frac{t_2\sin{k}}{\sqrt{t_1^2 + t_2^2 -2t_1t_2\cos{k}}}  \, \\
    &= \frac{1}{2} + \frac{\sin \theta \cos \phi}{2\pi} \int_{0}^{\pi} dk\,\frac{t_1-t_2\cos{k}}{\sqrt{t_1^2 + t_2^2 -2t_1t_2\cos{k}}}  \,,
\end{split}
\end{equation}
where we have used the fact that in the Brilluoin zone $k\in[-\pi,\pi]$, the second term involving $t_1 - t_2\cos{k}$ is an even function in $k$, whereas, the third term involving $t_2\sin{k}$ integrates to zero since it is an odd function in $k$.
Finally, we write the Krylov spread complexity by defining as follows,
\begin{align}\label{eq:appendixb_krylov_ssh}
    C(t_{1},t_{2}) = \frac{1}{2} + \Re(\alpha^*\beta)\mathcal{I}_{1}(t_{1},t_{2})\,. 
\end{align}
where we have defined the integral, 
\begin{align}\label{eq:ssh_krylov_integrals}
    \mathcal{I}_{1}(t_{1},t_{2}) = \frac{1}{\pi} \int_{0}^{\pi} dk \frac{t_1-t_2\cos{k}}{\sqrt{t_1^2 + t_2^2 -2t_1t_2\cos{k}}}\,.
\end{align}

\subsection{Formulas for elliptic integrals}\label{subsubsec:elliptic_formulas}
In this section, we are going to collect all the formulas that we are going to need in the upcoming analysis. All the formulas here are taken from already given in chapter $8$ of Gradshteyn and Ryzhik ($7^{\text{th}}$ edition)~\cite{gradshteyn2007table}, with the the transformation $x = k^{2}$. Variable $k$ is used in the Ref.~\cite{gradshteyn2007table}, but here we use $x$. The complete elliptic integral of the first, and second kind are defined
as,
\begin{align}
K(x) & =\int_{0}^{\pi/2}\frac{dk}{\sqrt{1-x\,\sin^{2}k}}\,, \qquad 
E(x) =\int_{0}^{\pi/2}dk\sqrt{1-x\,\sin^{2}k}\,.
\end{align}
The series expansion of $K(x)$, and $E(x)$ is given as follows,
\begin{align}\label{eq:exp_elliptic}
K(x) & =\frac{\pi}{2}\bigg[1+\frac{1}{4}\cdot x+\frac{9}{64}\cdot x^{2}+\cdots\bigg]\\
E(x) & =\frac{\pi}{2}\bigg[1-\frac{1}{4}\cdot x-\frac{3}{64}\cdot x^{2}+\cdots\bigg]
\end{align}
The function $K(x)$ has a logarithmic divergence as $x\to1$. By
defining,
\begin{equation}
x'=1-x
\end{equation}
 the leading order behavior is,
\begin{equation}
\lim_{x\to1}K(x)=\lim_{x'\to0^{+}}K(1-x')\to\frac{1}{2}\ln\frac{16}{x'}\qquad\text{as }x'\to0^{+}
\end{equation}
The leading order behavior of the function $E(x)$ as $x\to1$ or
$x'=1-x\to0$ is,
\begin{equation}
\lim_{x\to1}E(x)=\lim_{x'\to0^{+}}E(1-x')\to1
\end{equation}
Additionally, the function $K(x)$, and $E(x)$ have the following
relation,
\begin{equation}
\frac{dK(x)}{dx}=\frac{E(x)-(1-x)K(x)}{2x(1-x)}
\end{equation} 

\subsection{Momentum-dependent reference state}
In this subsection we establish a connection with the previous work by Caputa and Liu~\cite{PhysRevB.106.195125}. The reference state used in Ref.~\cite{PhysRevB.106.195125} has the following unit vector on the Bloch sphere,
\begin{align}\label{eq:ref_k_dep_state}
    \hat{n}_{\text{ref}}(k)=\begin{cases}
(0,0,+1) & \text{for }k\in[-\pi,0]\\
(0,0,-1) & \text{for }k\in[0,\pi]
\end{cases}
\end{align}
and hence, Eq.~\eqref{eq:ref_k_dep_state} tells us that the reference state is implicitly a momentum \textit{k-dependent} state. As a result, the Krylov complexity per momentum mode $\textit{k}$, using Eq.~\eqref{eq:C_k} is given by,
\begin{align}\label{Caputa_C_k}
    C_{k}(t_{1},t_{2})=\begin{cases}
\frac{1}{2} - \frac{t_2\sin{|k|}}{2\sqrt{t_1^2 + t_2^2 -2t_1t_2\cos{|k|}}} & \text{for }k\in[-\pi,0]\\
\frac{1}{2} - \frac{t_2\sin{k}}{2\sqrt{t_1^2 + t_2^2 -2t_1t_2\cos{k}}} & \text{for }k\in[0,\pi]
\end{cases}
\end{align}
It is clear from Eq.~\eqref{Caputa_C_k} that for every momentum pair $(k,-k)$, we have $C_k=C_{-k}$ with $k>0$. Hence, for positive and negative modes, Krylov complexity adds up. The integrated Krylov complexity in this case is therefore, given by the following,
\begin{align}\label{eq:cap_krylov}
    C_{\text{SSH}}(t_{1},t_{2})&= \frac{2}{2\pi}\bigg[\int_{0}^{\pi} dk \,\frac{1}{2}-\int_{0}^{\pi} dk \, \frac{t_2\sin{k}}{\sqrt{t_1^2 + t_2^2 -2t_1t_2\cos{k}}}\bigg] =\frac{1}{2}-\mathcal{I}_{3}(t_1,t_2)\,.
\end{align}
where,
\begin{align}
    \mathcal{I}_{3}(t_1,t_2)=\frac1\pi\int_0^\pi dk\,\frac{t_2\sin{k}}{\sqrt{t_1^2 + t_2^2 -2t_1t_2\cos{k}}}\,.
\end{align}
We now wish to explictly evaluate this integral $\mathcal{I}_{3}(t_1,t_2)$. 
Let $u=\cos k$ and this implies, $du=-\sin k\,dk$, and the limits of $k:0\to\pi$ implies $u:1\to-1$:
\begin{align}
\mathcal{I}_3(t_{1},t_{2})
&=\frac{t_2}{\pi}\int_{1}^{-1}\frac{-du}{\sqrt{t_1^2+t_2^2-2t_1t_2u}}
=\frac{t_2}{\pi}\int_{-1}^{1}\frac{du}{\sqrt{A-Bu}},
\end{align}
where $A=t_1^2+t_2^2$ and $B=2t_1t_2$.
Since $\int du/\sqrt{A-Bu}=-(2/B)\sqrt{A-Bu}$, we have,
\begin{align}
\mathcal{I}_3(t_{1},t_{2})
&=\frac{t_2}{\pi}\left[-\frac{2}{B}\sqrt{A-Bu}\right]_{-1}^{1}
=\frac{t_2}{\pi}\frac{1}{t_1t_2}\left(\sqrt{A+B}-\sqrt{A-B}\right).
\end{align}
Next, note the following,
\begin{equation}
\begin{split}
    \sqrt{A+B} &= \sqrt{t_1^2 + t_2^2 + 2t_1t_2} = t_1+t_2 \,,\\
    \sqrt{A-B} &= \sqrt{t_1^2 + t_2^2 - 2t_1t_2} = |t_{1} - t_{2}| \,.
\end{split}
\end{equation}
Hence, we obtain the following,
\begin{align}\label{eq:integral_I3}
    \mathcal{I}_3(t_{1},t_{2}) = \frac{t_{1}+t_{2}-|t_{1}-t_{2}|}{\pi\,t_{1}}\,.
\end{align}
Hence, the integrated Krylov complexity density is given by,
\begin{align}\label{eq:caputa_result}
   C_{\text{SSH}}(t_{1},t_{2}) = \frac{1}{2} - \frac{t_{1}+t_{2}-|t_{1}-t_{2}|}{\pi\,t_{1}}\,. 
\end{align}
and hence, we have non-analyticity in the Krylov spread complexity at $t_{1}=t_{2}$, which is the topological phase transition point. One should also note that the integral $\mathcal{I}_3(t_1,t_2)$ enters in the expression for $C(t_1,t_2)$, see Eq.~\eqref{eq:cap_krylov} only when one considers the implicit momentum $k$-dependent reference state as in Eq.~\eqref{eq:ref_k_dep_state}. For a $k$-independent reference state living on the Bloch sphere $\mathcal{I}_3(t_1,t_2)$ never enters in the expression of $C(t_1,t_2)$.

Next, using our formalism, we try to generalize the work by Caputa and Liu~\cite{PhysRevB.106.195125}. Instead of antipodal vectors on the Bloch sphere i.e., choosing a reference state as in Eq.~\eqref{eq:ref_k_dep_state}, one can choose a class of reference states where the corresponding vector on the Bloch sphere is given as follows,
\begin{align}\label{eq:ref_class_1}
    \hat{n}_{\text{ref}}(k)=\begin{cases}
(\sin \theta \cos\phi,\sin \theta \sin\phi,\cos \theta) & \text{for }k\in[-\pi,0]\\
(\sin \theta \cos\phi,\sin \theta \sin\phi,-\cos \theta) & \text{for }k\in[0,\pi]
\end{cases}
\end{align}
Our formalism does not need to know the exact form of the circuit Hamiltonian that is needed to evolve this reference state to the ground state of the SSH model. Using Eq.~\eqref{eq:C_k}, the Krylov complexity per momentum mode \textit{k} is given as,
\begin{equation}\label{eq:class_1}
    C_{k}(t_{1},t_{2})=\begin{cases}
\frac{1}{2} + \frac{\sin \theta \cos \phi}{2} \frac{t_1-t_2\cos{|k|}}{\sqrt{t_1^2 + t_2^2 -2t_1t_2\cos{|k|}}} - \frac{\cos \theta}{2} \frac{t_2\sin{|k|}}{\sqrt{t_1^2 + t_2^2 -2t_1t_2\cos{|k|}}} & \text{for }k\in[-\pi,0]\\[6pt]
\frac{1}{2} + \frac{\sin \theta \cos \phi}{2} \frac{t_1-t_2\cos{k}}{\sqrt{t_1^2 + t_2^2 -2t_1t_2\cos{k}}} - \frac{\cos \theta}{2} \frac{t_2\sin{k}}{\sqrt{t_1^2 + t_2^2 -2t_1t_2\cos{k}}} & \text{for }k\in[0,\pi] \\[6pt]
\end{cases}
\end{equation}
and hence, Eq.~\eqref{eq:class_1} implies $C_k=C_{-k}$, for every momentum pair $(k,-k)$ with $k>0$. For this class of reference states one has the following expression for the averaged Krylov spread complexity density,
\begin{align}
    C_{\text{SSH}}(t_1,t_2) = \frac{1}{2} - \frac{\cos \theta}{2}\mathcal{I}_3(t_1,t_2) + \frac{\sin \theta \cos \phi}{2}\mathcal{I}_1(t_1,t_2)\,.
\end{align}
Equivalently, expressing in terms of the ket $|\psi_{\text{ref}}\rangle$ in Eq.~\eqref{eq:ket_ref}, 
\begin{align}\label{eq:integrated complexity of SSH GS}
    C_{\text{SSH}}(t_1,t_2) = \frac{1}{2} - \frac{\Delta}{2}\mathcal{I}_3(t_1,t_2) + \Re(\alpha^*\beta)\mathcal{I}_2(t_1,t_2)\,, \qquad \Delta=|\beta|^2-|\alpha|^2\,, \quad \Re(\alpha^*\beta)=\frac{\sin \theta \cos \phi}{2}\,.
\end{align}

\subsection{Behavior of Krylov spread complexity  }
In this subsection, we present a thorough study of Krylov spread complexity by explicitly evaluating the integral $\mathcal{I}_1(t_1,t_2)$ and considering different forms of momentum \textit{k-independent} reference state. Let us now concentrate on the integral $\mathcal{I}_{1}(t_{1},t_{2})$:
\begin{align}\label{eq:intgeral_I1}
\mathcal{I}_{1}(t_{1},t_{2})=\frac1\pi\int_0^\pi dk\,\frac{t_1-t_2\cos k}{|\bold{d}_{\text{SSH}}(k)|}\,, \qquad |\bold{d}_{\text{SSH}}(k)| = \sqrt{t_1^2 + t_2^2 -2t_1t_2\cos{k}}\,.
\end{align}
Using $\cos k=1-2\sin^2(k/2)$ we simplify the denominator as follows,
\begin{align}
|\bold{d}_{\text{SSH}}(k)|^2
&=t_1^2+t_2^2-2t_1t_2\cos k
=(t_1-t_2)^2+4t_1t_2\sin^2\frac{k}{2}
=\delta^2+4t_1t_2\sin^2\frac{k}{2}\,.
\end{align}
Inorder to move forward, we define the following new variables that simplifies the calculations,
\begin{align}
s=t_1+t_2\,,\qquad \delta=t_1-t_2\,,\qquad
m=\frac{4t_1t_2}{(t_1+t_2)^2}=\frac{4t_1t_2}{s^2}\,,
\qquad
1-m=\left(\frac{\delta}{s}\right)^2\,.
\end{align}
This enables us to write the denominator, $d(k)$ in the following form,
\begin{align}
|\bold{d}_{\text{SSH}}(k)|^2
&=s^2\left((1-m)+m\sin^2\frac{k}{2}\right)
=s^2\left(1-m\cos^2\frac{k}{2}\right) \Rightarrow
d(k)=s\sqrt{1-m\cos^2\frac{k}{2}}.
\end{align}
Hence, the integral can be re-casted as:
\begin{align}
\mathcal{I}_{1}(t_{1},t_{2})
&=\frac1\pi\int_0^\pi dk\,
\frac{s-2t_2\cos^2\frac{k}{2}}{s\sqrt{1-m\cos^2\frac{k}{2}}}.
\end{align}
Let $\theta=k/2$ so $dk=2d\theta$ and $\theta\in[0,\pi/2]$:
\begin{align}
\mathcal{I}_{1}(t_{1},t_{2})
=\frac{2}{\pi}\int_0^{\pi/2} d\theta
\left[
\frac{1}{\sqrt{1-m\cos^2\theta}}
-\frac{2t_2}{s}\frac{\cos^2\theta}{\sqrt{1-m\cos^2\theta}}
\right].
\end{align}
By substituting $\varphi=\frac{\pi}{2}-\theta$ we obtain the complete elliptic integrals, cf. chapter $8$ of Gradshteyn and Ryzhik ($7^{\text{th}}$ edition)~\cite{gradshteyn2007table}, and also Appendix~\ref{subsubsec:elliptic_formulas},
\begin{equation}\label{eq:elliptic_integrals}
\begin{split}
\int_0^{\pi/2}\frac{d\theta}{\sqrt{1-m\cos^2\theta}}
&=\int_0^{\pi/2}\frac{d\varphi}{\sqrt{1-m\sin^2\varphi}}= K(m) \,,\\
\int_0^{\pi/2}\frac{\cos^2\theta\,d\theta}{\sqrt{1-m\cos^2\theta}}
&=\int_0^{\pi/2}\frac{\sin^2\varphi\,d\varphi}{\sqrt{1-m\sin^2\varphi}}
=\frac{K(m)-E(m)}{m}\,,
\end{split}
\end{equation}
where $K(m)$ is the complete elliptic integral of first kind and,
\begin{align}
E(m) = \int_0^{\pi/2} d\varphi\,\sqrt{1-m\sin^2\varphi},
\end{align}
is the complete elliptic integral of second kind, see integral $3.617$ in~\cite{gradshteyn2007table}, and also Appendix~\ref{subsubsec:elliptic_formulas}. Therefore, we obtain the following expression for the integral $\mathcal{I}_{1}(t_{1},t_{2})$,
\begin{align}
\mathcal{I}_{1}(t_{1},t_{2})
&=\frac{2}{\pi}\left[
K(m)-\frac{2t_2}{s}\frac{K(m)-E(m)}{m}
\right].
\end{align}
Using $m=4t_1t_2/s^2$ gives $\frac{2t_2}{s}\frac{1}{m}=\frac{s}{2t_1}$, we finally obtain:
\begin{align}\label{eq:integral_I1}
\mathcal{I}_1(t_1,t_2)=\frac{\delta\,K(m)+s\,E(m)}{\pi t_1},
\qquad
m=\frac{4t_1t_2}{(t_1+t_2)^2}.
\end{align}
As a result, the closed form expression of the Krylov complexity averaged over all momentum modes for the ground state of the SSH model starting from an arbitrary $k-$independent reference state,
\begin{align}\label{eq:kry_sprd_ssh_full}
C_{\text{SSH}}(t_1,t_2)
&=\frac12+\Re(\alpha^*\beta)\,\frac{\delta\,K(m)+s\,E(m)}{\pi t_1}\,,
\end{align}
where,
\begin{align}
    s = t_{1} + t_{2}\,, \qquad \delta = t_{1} - t_{2}\,, \qquad m=\frac{4t_1t_2}{(t_1+t_2)^2}\,.
\end{align}
Next, we would like to analyze $\mathcal{I}_{1}(t_{1},t_{2})$. Let us start by noting the following,
\begin{align}
    1-m=(\delta/s)^2\,, 
\end{align}
hence near the gap closing point we have $\delta \to 0$ and therefore $m\to1$. Note that the complete elliptic integral $K(m)$ have non-analyticity at $m=1$. From chapter $8$ of Gradshteyn and Ryzhik ($7^{\text{th}}$ edition), and also Appendix~\ref{subsubsec:elliptic_formulas} we obtain the asymptotic forms of the complete Elliptic integrals,
\begin{align}
    K(m \to 1) \approx \frac{1}{2}\ln \frac{16}{|1-m|} = \ln \frac{4\,s}{|\delta|}\,, \qquad E(m \to 1) \approx 1\,,
\end{align}
where we have neglected all the terms of $\mathcal{O}(\delta^{2}/s^{2})$. Therefore, we have the following form of the integral $\mathcal{I}_{1}(t_{1},t_{2})$ near the gap closing point i.e. $\delta/s \to 0$,
\begin{align}\label{eq:I1}
    \mathcal{I}_{1}(t_{1},t_{2}) \approx \frac{s}{\pi t_1}+\frac{\delta}{\pi t_1}\ln\frac{4\,s}{|\delta|}
+ \mathcal{O}(\delta^{2}/s^{2}) \,,
\end{align}
Now, let us come back to the expression for the Krylov spread complexity Eq.~\eqref{eq:kry_sprd_ssh_full} and we see that for a generic reference state (correspondingly the zeroth Krylov basis vector which is $k$ independent), Eq.~\eqref{eq:generic_ref_state} with $\alpha\,,\beta\,\neq 0$, only the integral $\mathcal{I}_{1}(t_{1},t_{2})$ contributes. As a result, the Krylov spread complexity in general cannot have plateau behavior in the topological regime. It is sensitive to the reference state we choose as we have shown here explicitly.

\section{Krylov Complexity of a general eigenstate of SSH}
\label{appendix:c}
In this section, we are going to generalize our analysis beyond the SSH ground state. By considering an excited state of the SSH Hamiltonian as the target state, we are going to evaluate the Krylov spread complexity. We find that the spread complexity in this case is also sensitive to the gap closing in the spectrum. Let us start by considering an excited state of the SSH Hamiltonian, see Eq.~\eqref{eq:ssh_ham_appen_a} as follows which is going to be our target state,
\begin{equation}
    \ket{\psi_{\text{target}}}
    =
    \otimes_k
    \ket{s(k)\mathbf E_k}\,,
    \label{eq:general_many_body_SSH_eigenstate_product}
\end{equation}
where the function $s(k)= \pm 1$ and determines whether for given $k$ we choose $\ket{+\mathbf E_k}$ or $\ket{-\mathbf E_k}$, cf. Eq.~\eqref{eq:GS vector in k-space}. Note that the state $\ket{\pm \mathbf E_k}$ are the eigenstates of the SSH Hamiltonian with eigenvalue $\pm \sqrt{t_{1}^{2}+t_{2}^{2}-2t_{1}t_{2}\cos k}$, see Appendix~\ref{appendix:a} for the details. In order to obtain the spread complexity, we choose reference state as $\ket{\psi}_{\text{ref}} =  \alpha \ket{\uparrow} + \beta \ket{\downarrow}$, see Eq.~\eqref{eq:arrow_kets_appen_b} and Eq.~\eqref{eq:ket_ref}. For each $k$, we can write the Bloch sphere vector corresponding to target state, and also for the reference state as follows using Eq.~\eqref{eq:groundstate_ssh_appen_b} and Eq.~\eqref{eq:generic_ref_state},
\begin{align}
    \hat{n}_{\text{target}}(k) &= s(k)\left(
\frac{t_1-t_2\cos{k}}{\sqrt{t_1^2 + t_2^2 -2t_1t_2\cos{k}}},
\,0,\,
\frac{t_2\sin{k}}{\sqrt{t_1^2 + t_2^2 -2t_1t_2\cos{k}}}
\right)\,, \\
\hat{n}_{\text{ref}}(k) &= (\cos \phi\sin \theta,\,\sin \phi\sin \theta,\,\cos \theta)\,.
\end{align}
For each momentum mode, we obtain the spread complexity as follows using Eq.~\eqref{eq:C_k}, 
\begin{align}
    C_k = \frac{1}{2} - \frac{s(k)}{2}\left[ \sin \theta \cos \phi \frac{t_1-t_2\cos{k}}{\sqrt{t_1^2 + t_2^2 -2t_1t_2\cos{k}}} + \cos \theta\frac{t_2\sin{k}}{\sqrt{t_1^2 + t_2^2 -2t_1t_2\cos{k}}} \right]\,.
\end{align}
Once, we have the expression for the spread complexity for each momentum mode $k$, let us consider a concrete example to evaluate the spread complexity. We start by partitioning the Brillouin zone as follows,
\begin{equation}
    -\pi=k_0<k_1<\cdots<k_N=\pi\,,
\end{equation}
and for a given interval $[k_i,\,k_{i+1}]$, we assume that for each $k\in [k_i,\,k_{i+1}]$ we have $s(k) = s_{k_i}$, and hence fixed for the entire interval. In other words, we choose $\ket{s_{k_{i}}\mathbf E_k}$ as our target state for each $k\in [k_i,\,k_{i+1}]$. Hence, we find that the spread complexity averaged over the entire Brillouin zone is given as follows,
\begin{equation}
    C(t_1,t_2)
     = \frac{1}{2\pi} \int_{-\pi}^{\pi} dk C_{k}(t_{1},t_{2}) =
    \frac{1}{2}
    +
    \frac{\sin \theta \cos \phi}{4\pi}
    \sum_{j=0}^{N-1}s_{k_j} \mathcal{I}_{1}(k_j,k_{j+1})
    +
    \frac{\cos \theta}{4\pi}
    \sum_{j=0}^{N-1}s_{k_{j}} \mathcal{I}_{3}(k_j,k_{j+1})\,,
    \label{eq:C_total_piecewise_general}
\end{equation}
where
\begin{align}
    \mathcal{I}_{1}(k_{j},k_{j+1})
    &=
    \int_{k_{j}}^{k_{j+1}} dk\,
    \frac{t_1-t_2\cos k}{\sqrt{t_{1}^{2} + t_{2}^{2} - 2t_{1}t_{2}\cos k}} = \int_{k_{j}}^{k_{j+1}} dk\, \frac{\partial}{\partial t_{1}} \sqrt{t_{1}^{2} + t_{2}^{2} - 2t_{1}t_{2}\cos k} = \frac{\partial}{\partial t_{1}} \int_{k_{j}}^{k_{j+1}} dk\, E_{k}\,,
    \label{eq:Ix_def_general}
    \\
    \mathcal{I}_{3}(k_{j},k_{j+1})
    &=
    \int_{k_{j}}^{k_{j+1}} dk\,
    \frac{t_2\sin k}{\sqrt{t_{1}^{2} + t_{2}^{2} - 2t_{1}t_{2}\cos k}} = \int_{k_{j}}^{k_{j+1}} dk\, \frac{d}{d k} \frac{\sqrt{t_{1}^{2} + t_{2}^{2} - 2t_{1}t_{2}\cos k}}{t_{1}} = \frac{E_{k_{j+1}} - E_{k_{j}}}{t_{1}}\,,
    \label{eq:Iy_def_general}
\end{align}
where we have used used $E_{k} = \sqrt{t_{1}^{2} + t_{2}^{2} - 2t_{1}t_{2}\cos k}$. The last part is to evaluate the integral $\mathcal{I}_{1}(k_{j},k_{j+1})$. It turns out that by writing $E_k$ as follows,
\begin{equation}
    E_k
    =
    (t_1+t_2)
    \sqrt{
        1-m\cos^2\frac{k}{2}
    }\,, \qquad m = \frac{4t_{1}t_{2}}{(t_{1}+t_{2})^{2}}\,,
    \label{eq:Ek_elliptic_form_general}
\end{equation}
we can identify $\mathcal{I}_1(k_j, k_{j+1})$ with an incomplete elliptic integral~\cite{gradshteyn2007table} denoted by $\mathbb{E}(a|b)$ as follows,
\begin{equation}
    \mathcal{I}_1(k_{j},k_{j+1})
    = \frac{\partial}{\partial t_{1}} \int_{k_{j}}^{k_{j+1}} dk\, E_{k} = 
    \frac{\partial}{\partial t_1}
    \left\{
        2(t_1+t_2)
        \left[
            \mathbb E\!\left(\frac{\pi-k_{j}}{2}\middle|m\right)
            -
            \mathbb E\!\left(\frac{\pi-k_{j+1}}{2}\middle|m\right)
        \right]
    \right\}.
    \label{eq:Ix_elliptic_general}
\end{equation}
\begin{figure}[t]
    \centering
    \includegraphics[width=0.95\linewidth]{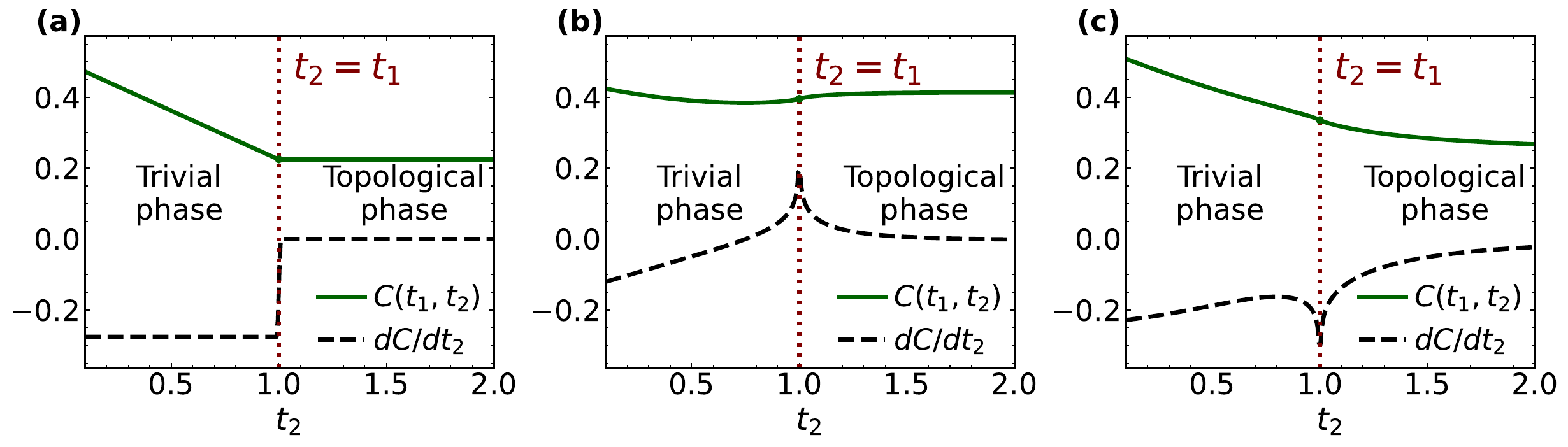}
    \caption{Krylov spread complexity $C(t_1,t_2)$ and its derivative $dC(t_1,t_2)/dt_2$ for the following two eigenstates, (a) Target state $\ket{\psi_{\text{target}}} = \otimes_{k \leq 0} \ket{\textbf{E}_k} \otimes_{k>0} \ket{-\textbf{E}_k}$, (b) target state $\ket{\psi_{\text{target}}} = \otimes_{k \leq -\pi/2} \ket{\textbf{E}_k} \otimes_{k>-\pi/2} \ket{-\textbf{E}_k}$, and (c) target state $\ket{\psi_{\text{target}}} = \otimes_{k \leq \pi/4} \ket{\textbf{E}_k} \otimes_{k>-\pi/4} \ket{-\textbf{E}_k}$. In all the three cases, the reference state, for all $k$-modes, is fixed to be $\ket{\psi_{\text{ref}}}_k = \cos(\frac{\pi}{12}) \ket{\uparrow} + e^{i \pi/3} \sin(\frac{\pi}{12}) \ket{\downarrow}$.}
    \label{fig:2 interval figures}
\end{figure}
Putting these results together, the complexity of an arbitrary piecewise eigenstate configuration is
\begin{equation}
    C(t_1,t_2)
    =
    \frac{1}{2}
    +
    \frac{\sin \theta \cos \phi}{4\pi}
    \sum_{j=0}^{N-1}
    s_{k_j} \mathcal{I}_1(a_j,a_{j+1})
    +
    \frac{\cos \theta}{4\pi t_1}
    \sum_{j=0}^{N-1}
    s_{k_j}
    \left(
        E_{a_{j+1}}-E_{a_j}
    \right)
\label{eq:C_final_piecewise_general}
\end{equation}
With the general expression for the spread complexity, let us consider some simple examples and evaluate the spread complexity using Eq.~\eqref{eq:C_total_piecewise_general}. We choose our target state as,
\begin{align}
    \ket{\psi_{\text{target}}} = \otimes_{k \leq k_{0}} \ket{\textbf{E}_k} \otimes_{k > k_{0}} \ket{-\textbf{E}_k}\,,
\end{align}
where $k_{0}$ is the momentum mode such that for $k\leq k_{0}$ we choose the eigenstate $\ket{\textbf{E}_k}$, and for $k>k_{0}$ we choose the other eigenstate $\ket{-\textbf{E}_k}$. The spread complexity is given as follows,
\begin{equation}
        C(t_1, t_2) = \frac{1}{2} + \frac{\sin \theta \cos \phi}{4 \pi} \left[\mathcal{I}_{1}(-\pi, k_0) - \mathcal{I}_{1}(k_{0}, \pi)\right] + \frac{\cos \theta}{4 \pi t_1} \left[(E_{k_0} - E_{-\pi}) - (E_{\pi} - E_{k_0})\right]\,.
    \end{equation}
From here, let us choose $k_{0} = 0$ and note that $E_{k}$ and $\partial_{t_{1}} E_{k}$ are even function of $k$, see Eq.~\eqref{eq:Ix_def_general}, and therefore $\mathcal{I}_{1}(-\pi, 0) = \mathcal{I}_{1}(0, \pi)$. Further we also have, $E_{-\pi} = E_{\pi} = t_{1} + t_{2}$, and $E_{0} = |t_{1}-t_{2}|$, note that in all our analysis we always assume $t_{1},\,t_{2} > 0$. This implies,
\begin{align}
    C(t_1, t_2) = \frac{1}{2}  + \frac{\cos \theta}{2 \pi t_1} \big{[}|t_{1}-t_{2}| - (t_{1}+t_{2})\big{]}\,,
\end{align}
and hence we have non-analyticity at the gap closing point $t_{1} = t_{2}$. For a general $k_{0} \neq 0$, we have the incomplete elliptic integrals to work with, and therefore we resort to numerics and plot both the complexity $C(t_1, t_2)$ and its derivative $dC(t_1, t_2)/dt_2$ to see the non-analytic behavior at the gap closing point $t_{1} = t_{2}$, see Fig.~\ref{fig:2 interval figures}. Thus, the non-analyticity of the complexity is not restricted to the ground state. More general eigenstate configurations can also detect the SSH transition: whenever the band assignment allows the gap-closing mode to contribute nontrivially, Eq.~\eqref{eq:C_final_piecewise_general} develops a singular feature at $t_1=t_2$. The reference state and the pattern of occupied bands control the amplitude of this signature, but not its location.

\section{Fidelity susceptibility and Krylov complexity for two-band Hamiltonians }\label{appendix:d}

We begin with a general two-band Hamiltonian with a parameter $\lambda$,
\begin{equation}\label{eq:fs_ham}
H(k, \lambda)=\pmb{\sigma} \cdot \textbf{d}(k, \lambda) = \sigma_{x}d_{x}(k, \lambda) + \sigma_{y}d_{y}(k, \lambda) + \sigma_{z}d_{z}(k, \lambda)  \,.
\end{equation}
In the Hamiltonian, we assume that by tuning $\lambda$, we go across the gap closing point. In this appendix, we study fidelity susceptibility~\cite{doi:10.1142/S0217979210056335}, which captures the gap closing in the Hamiltonian, Eq.~\eqref{eq:fs_ham}. Further, we also study Krylov spread complexity and its relation to fidelity susceptibility in two-band Hamiltonians.

\subsection{Fidelity Susceptibility for two-band models}

The ground state wavefunction of the Hamiltonian undergoes a significant change if there is gap closing point in the spectrum of the Hamiltonian. Fidelity susceptibility captures this gap closing point, which is defined using the ground state of the Hamiltonian, Eq.~\eqref{eq:fs_ham}. More explicitly, fidelity susceptibility is defined as follows,
\begin{equation}
\chi_F(k, \lambda)
=
\langle \partial_\lambda u_{-}(k, \lambda) \mid \partial_\lambda u_{-}(k, \lambda) \rangle
-
\left|\langle u_{-}(k, \lambda) \mid \partial_\lambda u_{-}(k, \lambda) \rangle\right|^2\,,
\label{eq:FS_def}
\end{equation}
where $u_{-}(k, \lambda)$ is the ground state of $H= \pmb{\sigma} \cdot \textbf{d}(k, \lambda)$, Eq.~\eqref{eq:fs_ham} with eigenvalue $-|\textbf{d}(k, \lambda)| = -\sqrt{d_{x}^{2}(k, \lambda) + d_{y}^{2}(k, \lambda) + d_{z}^{2}(k, \lambda)}$. A convenient way to evaluate this quantity is through the projector
onto the ground state $\ket{u_{-}(k,\lambda)}$. The fidelity susceptibility shows a divergence as we approach the gap closing point by tuning $\lambda$. Consider the projector onto the ground state,
\begin{equation}
P_-(k, \lambda)=|u_-(k, \lambda)\rangle\langle u_-(k, \lambda)|
=\frac{1}{2}\big[ \mathbb{I}_2- \pmb{\sigma}\cdot \hat{d}(k, \lambda)\big{]}\,,
\label{eq:fs_projector}
\end{equation}
where $\hat{d}(k, \lambda) = \textbf{d}(k, \lambda) / |\textbf{d}(k, \lambda)|$ and $\mathbb{I}_2$ is $2\times 2$ identity matrix..
Fidelity susceptibility can be written in terms of the ground state projector,
\begin{equation}
\chi_F(k,\lambda)
=
\frac{1}{2}\,\Tr\!\left\{[\partial_\lambda P_-(k, \lambda)]^2\right\}.
\label{eq:FS_projector}
\end{equation}
Next, we differentiate the ground state projector, Eq.~\eqref{eq:fs_projector} with respect to the parameter $\lambda$, and obtain the following,
\begin{equation}
\partial_\lambda P_- (k, \lambda)
=
-\frac{1}{2}\,\big{[}\partial_\lambda \hat{d}(k, \lambda)\big{]}\cdot \pmb{\sigma}\, \Rightarrow \, (\partial_\lambda P_- (k, \lambda))^2
=
\frac{1}{4}\big[(\partial_\lambda \hat{d}(k, \lambda)\cdot \pmb{\sigma}\big]^2\,.
\end{equation}
Next, we use the following vector identity,
\begin{equation}\label{eq:vec_id}
(\textbf{a}\cdot \pmb{\sigma})(\textbf{b}\cdot \pmb{\sigma})
=
(\textbf{a} \cdot \textbf{b})\,\mathbb{I}_2
+
i(\textbf{a}\times \textbf{b})\cdot \pmb{\sigma}\,,
\end{equation}
Hence, we have,
\begin{align}
    [\partial_\lambda P_-(k, \lambda)]^2
=
\frac{1}{4}\big[(\partial_\lambda \hat{d}(k, \lambda)\cdot \pmb{\sigma}\big]^2 = \frac{1}{4} |\partial_\lambda \hat{d}(k, \lambda)|^{2} \, \mathbb{I}_2 \,.
\end{align}
Hence, we obtain the fidelity susceptibility per momentum mode $k$ which we average over the full Brillouin zone to obtain fidelity susceptibility as,
\begin{align}
\chi_F(k,\lambda)
=
\frac{1}{2}\,\Tr\!\left\{[\partial_\lambda P_-(k, \lambda)]^2\right\} = \frac{1}{4} |\partial_\lambda \hat{d}(k, \lambda)|^{2} \, \Rightarrow \chi_F(\lambda) = \frac{1}{8\pi}\int_{-\pi}^\pi dk\, |\partial_\lambda \hat{d}(k, \lambda)|^{2} \,.
\label{eq:FS_final}
\end{align}
The quantity $\partial_\lambda \hat{d}(k, \lambda)$ is evaluated as follows,
\begin{align}
\partial_\lambda \hat d(k,\lambda)
=
\frac{1}{| \bold{d}(k,\lambda)|}
\left[
\partial_\lambda \bold{d}(k,\lambda)-\hat d(k,\lambda) \{ \hat d(k,\lambda) \cdot \partial_\lambda \bold{d}(k,\lambda) \}
\right]
\,,
\label{eq:derivative_unit_d_short}
\end{align}
which can be written compactly by defining a transverse projector $P_\perp(k,\lambda)$, which is the projector in the direction orthogonal to $\hat{d}(k,\lambda)$. We obtain,
\begin{equation} \label{eq:fidelity susceptibility per momentum mode}
    \chi_F(k, \lambda) = \frac{1}{4}  \frac{| P_\perp (k,\lambda) \partial_\lambda \bold{d}(k, \lambda)|^{2}}{|\bold{d}(k, \lambda)|^{2}} \,.
\end{equation}
Hence, from Eq.~\eqref{eq:fidelity susceptibility per momentum mode} we find that any non-analyticity in the fidelity susceptibility would come from the gap closing.

\subsection{Complexity and gap closing in a generic two-band Hamiltonian}
In this sub-section, we highlight the nature of complexity per momentum mode and its first derivative with respect to a tunable parameter $\lambda$, for a general two-band Hamiltonian. The Krylov spread complexity for the ground state of a two-band Hamiltonian is given as follows,
\begin{align}
C_{k}(\lambda)
=
\frac{1-\hat{n}_{\text{ref}}(k)\cdot [-\hat d(k,\lambda)]}{2}=\frac{1+\hat{n}_{\text{ref}}(k)\cdot\hat d(k, \lambda)}{2}\,,
\label{eq:Ck_def_proj}
\end{align}
where $\hat{n}_{\text{ref}}(k)$ is the Bloch vector corresponding to our choice of the reference state, and similarly $-\hat{d}(k,\lambda)$ is the corresponding Bloch vector for the ground state of the two-band Hamiltonian. From here, we are going to take the derivative fo the spread complexity to obtain,
\begin{align}
    \partial_{\lambda}C_{k}(\lambda)
= \frac{1}{2}\, \hat{n}_{\text{ref}}(k)\cdot \partial_{\lambda} \hat{d}(k,\lambda)\,.
\end{align}
Hence, using Eq.~\eqref{eq:derivative_unit_d_short}, we obtain the following expression for the derivative of the spread complexity,
\begin{align}
    \partial_{\lambda}C_{k}(\lambda)
= \frac{1}{2}\cdot \hat{n}_{\text{ref}}(k)\cdot \partial_{\lambda} \hat{d}(k,\lambda)= \frac{1}{2}\cdot \hat{n}_{\text{ref}}(k)\cdot \frac{1}{| \bold{d}(k,\lambda)|}
[P_\perp (k,\lambda) \partial_\lambda \bold{d} (k, \lambda)].
\end{align}
Therefore, just like in fidelity susceptibility, we find that any non-analyticity  in $\partial_{\lambda} C(\lambda)$ would come from the gap closing point.

\subsection{Relation between fidelity susceptibility and Krylov spread complexity}
Let us start with the fact that the Krylov spread complexity in two dimensional Hamiltonians of the form $H(k,\lambda) = \boldsymbol{\sigma} \cdot \mathbf{d(k,\lambda)}$ where $k$ is the momentum and $\lambda$ is a tunable parameter in the system is given as follows,
\begin{equation}
C(\lambda) = c_{0} + \sum_{i=1}^{3} C^{(i)}(\lambda) = c_{0} + \mathcal{Q}_{i}\cdot\int_{-\pi}^{\pi}dk\frac{d_{i}(k,\lambda)}{|\bold{d}(k,\lambda)|}=c_{0} + \mathcal{Q}_{i}\cdot\int_{-\pi}^{\pi}dk\cdot\hat{n}_{i}(k,\lambda)\,,
\end{equation}
where $c_{0}$ is constant number, cf. Eq.~\eqref{eq:appendixb_krylov_ssh} and $\hat{d}_{i}(k,\lambda) = d_{i}(k,\lambda)/|\bold{d}(k,\lambda)|$ with $|\bold{d}(k,\lambda)| = \sqrt{\sum_{i=1}^{3} d_{i}^{2}(k,\lambda)}$. From here we have,
\begin{align}
    \partial_{\lambda} C(\lambda) = \sum_{i=1}^{3} \partial_{\lambda} C^{(i)}(\lambda) = \sum_{i=1}^{3}\mathcal{Q}_{i}\,\int_{-\pi}^{\pi}dk\,\partial_{\lambda}\hat{d}_{i}(k,\lambda)\,.
\end{align}
Let us focus on single component, and by using Cauchy-Schwarz inequality, we find,
\begin{align}
\big[\partial_{\lambda}C^{(i)}(\lambda)\big]^{2} & =\mathcal{Q}_{i}^{2}\Big[\int_{-\pi}^{\pi}dk\,\partial_{\lambda}\hat{d}_{i}(k,\lambda)\Big]^{2} =\mathcal{Q}_{i}^{2}\Big[\int_{-\pi}^{\pi}dk\cdot1\cdot\partial_{\lambda}\hat{d}_{i}(k,\lambda)\Big]^{2}\leq 2\pi\mathcal{Q}_{i}^{2}\int_{-\pi}^{\pi}dk\,\big[\partial_{\lambda}\hat{d}_{i}(k,\lambda)\big]^{2}\,.
\end{align}
Next, by using the expression for fidelity susceptibility and we get,
\begin{equation}
\frac{\big[\partial_{\lambda}C^{(i)}(\lambda)\big]^{2}}{2\pi\mathcal{Q}_{i}^{2}}\leq\int_{-\pi}^{\pi}dk\,\big[\partial_{\lambda}\hat{d}_{i}(k,\lambda)\big]^{2}=8\pi\chi_{F}^{i}(\lambda)\,, \qquad \chi_{F}^{i}(\lambda) = \frac{1}{2\pi} \cdot \frac{1}{4} \int_{-\pi}^{\pi} dk\,|\partial_{\lambda} \hat{d}_{i}(k,\lambda)|^2\,,
\end{equation}
or,
\begin{align}\label{eq:component_fide_kry_inequality}
|\partial_{\lambda}C^{(i)}(\lambda)| & \leq \, 4\pi|\mathcal{Q}_{i}|\,\sqrt{\chi_{F}^{i}(\lambda)}\,, \qquad \text{for }i=1,2,3\,.
\end{align}
From here, we can use the triangle inequality to obtain the following,
\begin{align}
    |\partial_{\lambda} C(\lambda)| = \left| \sum_{i=1}^{3} \partial_{\lambda}C^{(i)}(\lambda) \right| \leq \sum_{i=1}^{3} \left| \partial_{\lambda}C^{(i)}(\lambda) \right|\,.
\end{align}
Hence by using Eq.~\eqref{eq:component_fide_kry_inequality}, we obtain the following inequality for the Krylov spread complexity,
\begin{align}\label{eq:fs_component}
    |\partial_{\lambda} C(\lambda)| \leq 4\pi\,\sum_{i=1}^{3} |\mathcal{Q}_{i}|\,\sqrt{\chi_{F}^{i}(\lambda)}\,.
\end{align}
We would like to emphasize that the above inequality Eq.~\eqref{eq:fs_component} holds when the reference state (information is absorbed inside $|\mathcal{Q}_{i}|$) is momentum $k-$independent.

\subsection{Fidelity susceptibility and Krylov spread complexity in the massive Dirac Hamiltonian }

In this subsection, we explicitly show the inequality between the first derivative of the Krylov spread complexity and fidelity susceptibility, Eq.~\eqref{eq:component_fide_kry_inequality}. Consider the massive Dirac Hamiltonian, Eq.~\eqref{eq:md_ham_k},
\begin{equation}\label{Eq:massive_Dirac}
H_{\text{MD}}(k,\mu) = \textbf{d}_{\text{MD}}(k,\mu) \cdot \pmb{\sigma} = \sin k\,\sigma_x + \mu\,\sigma_z,
\qquad k\in[-\pi,\pi]\,.
\end{equation}
Two eigenvalues of the massive Dirac Hamiltonian is given as follows,
\begin{equation}
E_{\pm}(k)=\pm \sqrt{\sin^2 k+\mu^2}\,,
\end{equation}
and hence the gap closes at $\mu=0$ and $k=0,\pi$. As demonstrated previously in Appendix~\ref{subsubsec:massive_dirac_winding}, there is no topological phase transition across the gap closing point. The Krylov spread complexity averaged over the Brillouin zone as a function of the tunable parameter $\mu$ is,
\begin{align}
C_{\text{MD}}(\mu)&=\frac{1}{2}+\frac{\sin{\theta}\cos{\phi}}{4\pi}\int_{-\pi}^{\pi}\frac{\sin{k}}{\sqrt{\sin^2 k+\mu^2}}\,dk+\frac{\cos{\theta}}{4\pi}\int_{-\pi}^{\pi}\frac{\mu}{\sqrt{\sin^2 k+\mu^2}}\,dk\,,\\
&= \frac{1}{2} + C_{\text{MD}}^{(1)} + C_{\text{MD}}^{(2)} \,,
\end{align}
where $C_{\text{MD}}^{(1)}$, and $C_{\text{MD}}^{(2)}$ are the $x,\,z$ components of the Krylov spread complexity. Hence, the Krylov spread complexity is given as follows,
\begin{align}
    C_{\text{MD}}(\mu)=\frac{1}{2}+\frac{\cos{\theta}}{4\pi}\int_{-\pi}^{\pi}\frac{\mu}{\sqrt{\sin^2 k+\mu^2}}\,dk = \frac{1}{2} + \frac{\cos{\theta}}{4\pi} \cdot \mathcal{I}(\mu)\,,\label{eq:massive_dirac_krylov_complexity}
\end{align}
because the second integral vanishes as the integrand is an odd function of $k$ and $\mathcal{I}(\mu)$ is,
\begin{align}
    \mathcal{I}(\mu) = \int_{-\pi}^{\pi}\frac{\mu}{\sqrt{\sin^2 k+\mu^2}}\,dk\,.
\end{align}
Next, we use the fact the integrand is even function of $k$ and we can simplify the integral as follows,
\begin{align}
    \mathcal{I}(\mu) =& 2\mu\int_{0}^{\pi}\frac{1}{\sqrt{\sin^2 k+\mu^2}}\,dk = 2\mu\int_{0}^{\pi/2}\frac{1}{\sqrt{\sin^2 k+\mu^2}}\,dk + 2\mu\int_{\pi/2}^{\pi}\frac{1}{\sqrt{\sin^2 k+\mu^2}}\,dk \,,\\
    =& 2\mu\int_{0}^{\pi/2}\frac{1}{\sqrt{\sin^2 k+\mu^2}}\,dk + 2\mu\int_{0}^{\pi/2}\frac{1}{\sqrt{\cos^2 k+\mu^2}}\,dk \,, \\
    =& 4\mu\int_{0}^{\pi/2}\frac{1}{\sqrt{\sin^2 k+\mu^2}}\,dk\,.
\end{align}
From here, we identify this with the complete elliptic integral of the first kind, cf. Appendix~\ref{subsubsec:elliptic_formulas}, Eq.~\eqref{eq:elliptic_integrals}, and Ref.~\cite{gradshteyn2007table} as follows,
\begin{align}
    \mathcal{I}(\mu)= 4\mu\int_{0}^{\pi/2}\frac{1}{\sqrt{\sin^2 k+\mu^2}}\,dk\, = \frac{2\mu}{\sqrt{1+\mu^{2}}}\cdot K\left(\frac{1}{1+\mu^{2}}\right)\,.
\end{align}
Therefore, the integrated Krylov complexity, Eq.~\eqref{eq:massive_dirac_krylov_complexity} is given as follows,
\begin{equation}
C_{\text{MD}}(\mu)= \frac{1}{2}+\frac{\mu\cos\theta}{\pi\sqrt{1+\mu^2}}
K\!\left(\frac{1}{1+\mu^2}\right)\,.
\end{equation}
The complexity $C(\mu)$ is equal to $\frac{1}{2}$, for any reference state with $\cos{\theta}=0$. Next, we calculate the first derivative of the Krylov spread complexity with respect to $\mu$, and find its behavior near the gap closing point $\mu=0$. 
Let us start by defining,
\begin{equation}
\lambda=\frac{1}{1+\mu^2}\,,
\end{equation}
and differentiating with respect to $\mu$, we get,
\begin{equation}
\partial_{\mu}C_{\text{MD}}(\mu)=
\frac{\cos\theta}{\pi}
\left[
\frac{d}{d\mu}\bigl(\mu(1+\mu^2)^{-1/2}\bigr)K(\lambda)
+
\mu(1+\mu^2)^{-1/2}\frac{dK(\lambda)}{d\lambda}\frac{d\lambda}{d\mu}
\right].
\end{equation}
We now use the standard identity from Gradshteyn and Ryzhik~\cite{gradshteyn2007table}, chapter 8, Eq.~(8.123), and Appendix~\ref{subsubsec:elliptic_formulas},
\begin{equation}
\frac{dK(\lambda)}{d\lambda}
= \frac{E(\lambda)-(1-\lambda)K(\lambda)}{2\,\lambda\,(1-\lambda)}\,,
\end{equation}
This leads us to the following,
\begin{align}
\partial_{\mu}C_{\text{MD}}(\mu) &=
\frac{\cos\theta}{\pi}
\left[
\frac{K(\lambda)}{(1+\mu^2)^{3/2}}
-
\frac{\mu}{\sqrt{1+\mu^2}}
\frac{E(\lambda)-(1-\lambda)K(\lambda)}{2\,\lambda\,(1-\lambda)}\frac{2\mu}{(1+\mu^2)^2}
\right]\,,\\
&= 
\frac{\cos\theta}{\pi}
\left[
\frac{K(\lambda)}{(1+\mu^2)^{3/2}}
-
\frac{2\mu^{2}}{(1+\mu^2)^{5/2}}
\frac{(1+\mu^{2})^{2}E(\lambda)}{2\mu^{2}} + \frac{2\mu^{2}}{(1+\mu^2)^{5/2}}\frac{(1+\mu^{2})}{2}\,K(\lambda)
\right]\,,\\
&=\frac{\cos\theta}{\pi\,\sqrt{1+\mu^{2}}}
\left[
K(\lambda) - 
E(\lambda) 
\right]\,.\label{eq:derivative_krylov_massive_dirac}
\end{align}
Next, we need to evaluate the limit of $\mu\to 0$, this leads to the following,
\begin{align}\label{eq:derivative_c_mu_to_0}
    \lim_{\mu \to 0}\partial_{\mu}C_{\text{MD}}(\mu) &= \frac{\cos\theta}{\pi\,\sqrt{1+\mu^{2}}}
\left[
K(\lambda) - 
E(\lambda) 
\right]\,, \\
&= \lim_{\mu \to 0}\frac{\cos\theta}{\pi}
\left[
K\left(\frac{1}{1+\mu^{2}}\right) - E\left(\frac{1}{1+\mu^{2}}\right) 
\right]\,,\\
&= \frac{\cos\theta}{\pi}\left[ \ln \frac{4}{\mu} - 1 \right]\,,
\end{align}
and hence, a logarithmic divergence in the derivative of the Krylov spread complexity near the gap closing point,
\begin{align}\label{eq:lim_mu_0_krylov_massive_dirac}
     \lim_{\mu \to 0}\partial_{\mu}C_{\text{MD}}(\mu) \sim \frac{\cos\theta}{\pi}
\cdot \ln \frac{4}{|\mu|}\,.
\end{align}
This implies that Krylov spread complexity is not only sensitive to topological phase transitions but also gap closing points in the Hamiltonian, which in general can be a trivial system with no topological phase transition. 

Next, we evaluate the fidelity susceptibility for the massive Dirac Hamiltonian. Consider the following,
\begin{align}
    \hat d_{\text{MD}}(k,\mu)=\frac{ \textbf{d}_{\text{MD}}(k,\mu)}{|\textbf{d}_{\text{MD}}(k,\mu)|} =
-\left(
\frac{\sin k}{\sqrt{\sin^{2}k + \mu^{2}}},
\,0,\,
\frac{\mu}{\sqrt{\sin^{2}k + \mu^{2}}}
\right)\,.
\end{align}
From here, we can use the definition of the fidelity susceptibility, cf. Eq.~\eqref{eq:FS_final},
\begin{align}
    \chi_F^{\text{MD}}(k,\mu)=\frac{1}{4}\,\bigl|\partial_\mu \hat{d}_{\text{MD}}(k,\mu)\bigr|^2\,.
\end{align}
Let us next evaluate $\partial_\mu \hat{d}_{\text{MD}}(k,\mu)$,
\begin{align}
    \partial_\mu \hat{d}_{\text{MD}}(k,\mu) &= -\left(
-\frac{\mu \,\sin k}{(\sin^{2}k + \mu^{2})^{3/2}},
\,0,\,
\frac{1}{\sqrt{\sin^{2}k + \mu^{2}}} -\frac{\mu^{2}}{(\sin^{2}k + \mu^{2})^{3/2}}
\right) \,,\\
& = -\left(
-\frac{\mu \,\sin k}{(\sin^{2}k + \mu^{2})^{3/2}},
\,0,\,\frac{\sin^{2}k}{(\sin^{2}k + \mu^{2})^{3/2}}
\right)\,\Rightarrow \bigl|\partial_\mu \hat{d}_{\text{MD}}(k,\mu)\bigr|^2 = \frac{\sin^{2} k}{(\sin^{2} k + \mu^{2})^{2}}\,,\label{eq:del_d_massive_dirac}
\end{align}
and hence, we find,
\begin{align}\label{eq:massive_dirac_fs}
\chi_{F}^{\text{MD}}(\mu) =\frac{1}{2\pi}\frac{1}{4}\int_{-\pi}^{\pi} dk \bigl|\partial_\mu \hat{d}_{\text{MD}}(k,\mu)\bigr|^2  =\frac{1}{8\pi}\int_{-\pi}^{\pi} dk \,\frac{\sin^{2} k}{(\sin^{2} k + \mu^{2})^{2}}  = \frac{1}{8|\mu|(1+\mu^{2})^{3/2}} \,.
\end{align}
In the following we evaluate the integral explicitly, where a part of the result were obtained using Wolfram Mathematica. Consider the integral,
\begin{align}\label{eq:wolf_massive_dirac}
    \mathcal{I}(a)=\int_{0}^{\pi/2}\frac{dk}{\mu^{2}+a\sin^{2}k}=\frac{\pi}{2|\mu|\sqrt{\mu^{2}+a}}\,,
\end{align}
which is obtained using Wolfram Mathematica. Next consider the derivative with respect to the variable $a$,
\begin{align}
    \frac{\partial}{\partial a}\mathcal{I}(a)=-\int_{0}^{\pi/2}dk\frac{\sin^{2}k}{(\mu^{2}+a\sin^{2}k)^{2}}=-\frac{\pi}{2|\mu|}\cdot\frac{1}{2}\frac{1}{(\mu^{2}+a)^{3/2}}=-\frac{\pi}{4|\mu|(\mu^{2}+a)^{3/2}}\,,
\end{align}
and we find,
\begin{align}
    -\frac{\partial}{\partial a}\mathcal{I}(a)\Big|_{a=1}=\int_{0}^{\pi/2}dk\frac{\sin^{2}k}{(\mu^{2}+\sin^{2}k)^{2}}=\frac{\pi}{4|\mu|(\mu^{2}+1)^{3/2}}\,.
\end{align}
Next, consider the following to arrive at exact expression for the Fidelity susceptibility, Eq.~\eqref{eq:massive_dirac_fs},
\begin{align}\label{eq:integral_mani_massive_dirac}
    \int_{-\pi}^{\pi} dk \,\frac{\sin^{2} k}{(\sin^{2} k + \mu^{2})^{2}} &= \int_{0}^{\pi} dk \,\frac{2\sin^{2} k}{(\sin^{2} k + \mu^{2})^{2}} = \int_{0}^{\pi/2} dk \,\frac{2\sin^{2} k}{(\sin^{2} k + \mu^{2})^{2}}+\int_{\pi/2}^{\pi} dk \,\frac{2\sin^{2} k}{(\sin^{2} k + \mu^{2})^{2}}\,,\\
    &=\int_{0}^{\pi/2} dk \,\frac{2\sin^{2} k}{(\sin^{2} k + \mu^{2})^{2}}+\int_{0}^{\pi/2} dk \,\frac{2\cos^{2} k}{(\cos^{2} k + \mu^{2})^{2}}\,,\\
    &=\int_{0}^{\pi/2} dk \,\frac{4\sin^{2} k}{(\sin^{2} k + \mu^{2})^{2}}\,.
\end{align}
Next, let us compile everything up, and we have,
\begin{align}
    \chi_{F}^{\text{MD}}(\mu) &=\frac{1}{8\pi}\int_{-\pi}^{\pi} dk \,\frac{\sin^{2} k}{(\sin^{2} k + \mu^{2})^{2}} =  \frac{1}{2\pi}\int_{0}^{\pi/2} dk \,\frac{\sin^{2} k}{(\sin^{2} k + \mu^{2})^{2}} = \frac{1}{8|\mu|(\mu^{2}+1)^{3/2}}\,.
\end{align}
Hence, we obtain the full expression for the fidelity susceptibility for massive Dirac Hamiltonian, Eq.~\eqref{eq:massive_dirac_fs}. Near the gap closing point ($\mu\to 0$) we have a divergence as expected for the massive Dirac Hamiltonian,
\begin{align}
    \lim_{\mu \to 0} \chi_{F}^{\text{MD}}(\mu) \approx \frac{1}{8|\mu|}\,,
\end{align}
Finally, we would like verify the inequality between fidelity susceptibility and Krylov spread complexity, Eq.~\eqref{eq:component_fide_kry_inequality}. Therefore, let us start with considering the third component of the fidelity susceptibility $\chi_F^{\text{MD}(3)}(\mu)$, obtained from Eq.~\eqref{eq:del_d_massive_dirac},
\begin{align}\label{eq:third_comp_fs_massive_dirac}
    \chi_F^{\text{MD}(3)}(\mu) =\frac{1}{2\pi}\frac{1}{4}\int_{-\pi}^{\pi} dk\,\bigl|\partial_\mu \hat{d}_{\text{MD}}^{(3)}(k,\mu)\bigr|^2 = 
\frac{1}{8\pi}\int_{-\pi}^{\pi}\frac{\sin^4 k}{(\sin^2 k+\mu^2)^3}\,dk = \frac{3}{32|\mu| (1 + \mu^{2})^{5/2}}\,.
\end{align}
The integral is evaluated by using Eq.~\eqref{eq:wolf_massive_dirac}, Eq.~\eqref{eq:integral_mani_massive_dirac}, and the following,
\begin{align}
    \frac{1}{2}\frac{\partial^{2}}{\partial a^{2}}\mathcal{I}(a)\Big|_{a=1}&=\int_{0}^{\pi/2}dk\frac{\sin^{4}k}{(\mu^{2}+\sin^{2}k)^{2}}=\frac{3\pi}{16|\mu|(1+\mu^{2})^{5/2}}\,.
\end{align}
Hence, for massive Dirac Hamiltonian, we have Eq.~\eqref{eq:lim_mu_0_krylov_massive_dirac} for third component of Krylov spread complexity and Eq.~\eqref{eq:third_comp_fs_massive_dirac} for third component for fidelity susceptibility,
\begin{align}
    \lim_{\mu \to 0}\partial_{\mu} C_{\text{MD}}(\mu) = \partial_{\mu} C_{\text{MD}}^{(3)}(\mu) \sim \ln \frac{1}{|\mu|} \,, \qquad \lim_{\mu \to 0} \sqrt{\chi_F^{\text{MD}(3)}(\mu)} \sim \frac{1}{\sqrt{|\mu|}}\,,
\end{align}
near the gap closing point. We can clearly see that the divergence in the fidelity susceptibility is power law, whereas for the derivative of the Krylov spread complexity it is logarithmic and hence, a verification of the inequality, Eq.~\eqref{eq:FS_final} near the gap closing point $\mu=0$. Finally, we would also like to understand the behavior of the fidelity susceptibility and spread complexity in the limit of large $\mu$. Let us start with the Eq.~\eqref{eq:fs_component},
\begin{align}
    |\partial_{\lambda} C(\lambda)| \leq 4\pi\,\sum_{i=1}^{3} |\mathcal{Q}_{i}|\,\sqrt{\chi_{F}^{i}(\lambda)}\,.
\end{align}
Since, for massive Dirac Hamiltonian, only third component contributes for derivative of the Krylov spread complexity Eq.~\eqref{eq:derivative_krylov_massive_dirac}, therefore we also look at the third component of fidelity susceptibility,
\begin{align}
  \partial_{\mu}C_{\text{MD}}^{(3)}(\mu)  &=\frac{\cos\theta}{\pi\,\sqrt{1+\mu^{2}}}
\left[
K(\lambda) - 
E(\lambda) 
\right] \,, \qquad \lambda = \frac{1}{1+\mu^{2}}\,,\\
4\pi |\mathcal{Q}_{3}| \sqrt{\chi_F^{\text{MD}(3)}(\mu)}&=  |\cos\theta | \, \sqrt{\frac{3}{32|\mu| (1 + \mu^{2})^{5/2}}}\,, \qquad \mathcal{Q}_{3} = \frac{\cos \theta}{4\pi}\,.
\end{align}
In the limit of large $\mu$, we have, cf. Appendix~\ref{subsubsec:elliptic_formulas}
\begin{align}\label{eq:verify_inq}
    \frac{1}{\sqrt{1+\mu^{2}}}\left[
K\left(\frac{1}{1+\mu^{2}}\right) - 
E\left(\frac{1}{1+\mu^{2}}\right)
\right] \to \lim_{\mu\to\infty } \frac{\pi}{2|\mu|}\left[
1 +\frac{1}{4(1+\mu^{2})} - 1 + \frac{1}{4(1+\mu^{2})}
\right] \approx \frac{\pi}{4|\mu|^{3}}\,.
\end{align}
Using the above Eq.~\eqref{eq:verify_inq}, we get the following for the derivative of Krylov spread complexity and fidelity susceptibility,
\begin{align}
    \partial_{\mu}C_{\text{MD}}^{(3)}(\mu) &\approx \frac{\cos\theta}{4|\mu|^{3}} \,,\\
    4\pi |\mathcal{Q}_{3}| \sqrt{\chi_F^{\text{MD}(3)}(\mu)} &\approx \sqrt{\frac{3}{2}} \frac{ |\cos\theta|}{4|\mu|^{3}}\,.
\end{align}
Finally, we define the following ratio,
\begin{align}
    R(\mu) = \frac{|\partial_{\mu}C_{\text{MD}}^{(3)}(\mu)|}{4\pi |\mathcal{Q}_{3}| \sqrt{\chi_F^{\text{MD}(3)}(\mu)}}\,,
\end{align}
In the limit of large $\mu$, we get,
\begin{align}\label{eq:ratio_MD}
    \lim_{\mu\to\infty} R(\mu) = \lim_{\mu\to\infty} \frac{|\partial_{\mu}C_{\text{MD}}^{(3)}(\mu)|}{4\pi |\mathcal{Q}_{3}| \sqrt{\chi_F^{\text{MD}(3)}(\mu)}} = \sqrt{\frac{2}{3}}<1\,.
\end{align}

\subsection{Cooper pair box}
It is interesting to note that the Hamiltonian in Eq.~\eqref{Eq:massive_Dirac} can be realized in a physical setup such as the Cooper pair box.
\begin{figure}[h!]
    \centering
    \includegraphics[width=0.5\linewidth]{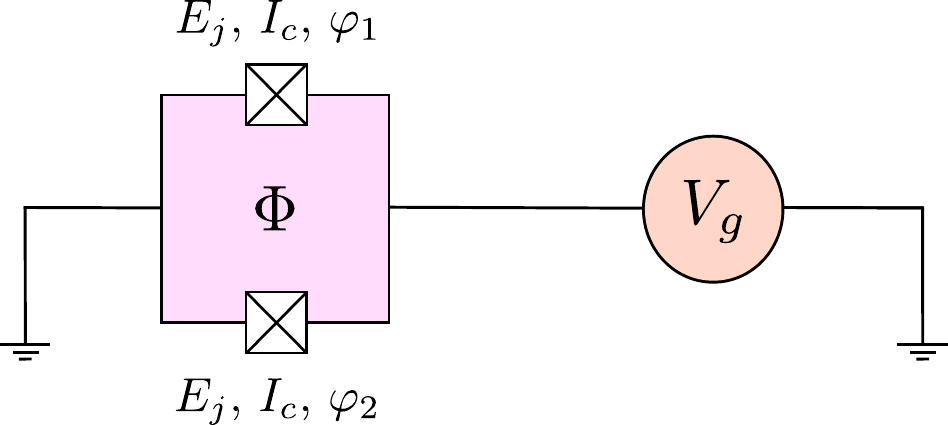}
    \caption{Illustration of a Cooper pair box, which realizes the physics of the massive Dirac Hamiltonian. The two Josephson junctions have identical Josephson junction energy $E_j$ and critical current $I_c$, and with different phases $\varphi_1$ and $\varphi_2$. The external gate voltage $V_g$ controls charging energy of the system. The magnetic flux $\Phi$ threading through the region between the Josephson junctions is a periodic parameter, which mimics the role of momentum in the Hamiltonian.}
    \label{fig:Josephson_junction}
\end{figure}
A Cooper pair box consists of two Josephson junctions
with a flux $\Phi$ threading through the region between the two Joshepson
junctions~\cite{Wong2025QuantumComputingArchitecture}. Further, we also choose the two Josephson junctions to be identical
i.e., same Josephson junction energy $E_{j}$ and identical critical
current $I_{c}$, see Fig.~\ref{fig:Josephson_junction}. In this setup with two identical Josephson junction
with flux $\Phi$ threading, the effective Josephson junction energy
depends on $\Phi$ with periodicity $2n\Phi_{0}$, where $n$ is an
integer, and $\Phi_{0}=h/2e$ is the flux quantum. The effective
Josephson junction energy is given as follows,
\begin{equation}
E_{j}^{\text{eff}}(\Phi)=2E_{j}\cos\Big(\frac{\pi\Phi}{\Phi_{0}}\Big)\,,
\end{equation}
The voltage source $V_{g}$, controls the effective energy $E=E_{cc}(1-2n_{g})$, where
$E_{cc}=2e^{2}/C$, is the charging energy of a Cooper pair, $C$ is the effective capacitance
in the system and $n_{g}$ is the number of extra Cooper pairs due
to external voltage source $V_{g}$. The full Hamiltonian can be described by a two-level system and is given
as follows,
\begin{equation}
H=-\frac{E_{j}^{\text{eff}}(\Phi)}{2}\sigma_{x}-\frac{E}{2}\sigma_{z}=\mathbf{d}(\Phi)\cdot\sigma\,,
\end{equation}
where we have defined,
\begin{equation}
\mathbf{d}(\Phi)=\bigg(-\frac{E_{j}^{\text{eff}}(\Phi)}{2},0,-\frac{E}{2}\bigg)\,.
\end{equation}
From here we find that there is gap closing as we tune $n_{g}=\frac{1}{2}$  and with $\Phi=\Phi_{0}/2$. The physics here is identical with
the massive Dirac Hamiltonian [cf. Eq.~\eqref{Eq:massive_Dirac}], where we can make the following identification,
\begin{align}
\mu & \equiv-\frac{E}{2}=e^{2}(1-2n_{g})/C\,,\\
\sin k & \equiv-\frac{E_{j}^{\text{eff}}(\Phi)}{2}=-E_{j}\cos\Big(\frac{\pi\Phi}{\Phi_{0}}\Big)\,.
\end{align}
 and hence, in Cooper pair box, the flux $\Phi$ mimics the role of
the momentum $k$ in the massive Dirac Hamiltonian .

\subsection{Fidelity susceptibility and Krylov spread complexity in the SSH model } 

In this subsection, we explicitly show that the inequality between the derivative of the spread complexity and the fidelity susceptibility, Eq.~\eqref{eq:component_fide_kry_inequality} is satisfied for the SSH model. Note that for the SSH model, the only non-vanishing contribution was from $\partial_{t_2}C_{\text{SSH}}^{(1)}(t_2)$ where the superscript $1$ refers to the $x-$component. Hence, we only need to evaluate $\partial_{t_2}C_{\text{SSH}}^{(1)}(t_2)$ and $\chi_{F}^{\text{SSH}(1)}(t_2)$ and check the inequality Eq.~\eqref{eq:component_fide_kry_inequality}. We now proceed to evaluate the first component of the fidelity susceptibility for the SSH model. For the SSH model, we have the following,
\begin{align}
    \hat d_{\text{SSH}}(k,t_1,t_2) =
-\left(
\frac{t_1-t_2\cos{k}}{\sqrt{t_1^2 + t_2^2 -2t_1t_2\cos{k}}},
\,0,\,
\frac{t_2\sin{k}}{\sqrt{t_1^2 + t_2^2 -2t_1t_2\cos{k}}}
\right)\,.
\end{align}
From this we can evaluate $\partial_{t_2} \hat{d}_{\text{SSH}}(k,t_1,t_2)$,
\begin{align}
    \partial_{t_2} \hat{d}_{\text{SSH}}(k,t_1,t_2) = -\left(
\frac{-t_1t_2\sin^2{k}}{(t_1^2 + t_2^2 -2t_1t_2\cos{k})^\frac{3}{2}},
\,0,\,
\frac{t_1\sin{k}(t_1-t_2\cos{k})}{(t_1^2 + t_2^2 -2t_1t_2\cos{k})^\frac{3}{2}}
\right)\,.
\end{align}
Hence, the first component of fidelity susceptibility is given by the following expression,
\begin{align}\label{eq:FS3_for_SSH}
 \chi_F^{\text{SSH}(1)}(t_2) =\frac{1}{2\pi}\frac{1}{4}\int_{-\pi}^{\pi} dk\,\bigl|\partial_{t_2} \hat{d}_{\text{SSH}}^{(1)}(k,t_2)\bigr|^2\,.  
\end{align}
Note that in Eq.\eqref{eq:FS3_for_SSH}, we tune the parameter $t_2$ in order to go across the gap closing point which in this case also corresponds to the topological phase transition point. Next, note the following integral identity obtained using Wolfram Mathematica,
\begin{align}\label{eq:Idenitity}
    \int_{-\pi}^{\pi}
\frac{\sin^4 k}{(A - B\cos k)^3}\,dk
=
\frac{3\pi(A - \Delta)^2}{B^4\Delta}\,, \qquad \Delta=\sqrt{A^2-B^2}\,,
\end{align}
and using Eq.~\eqref{eq:Idenitity}, we obtain the following,
\begin{align}
\chi_F^{\text{SSH}(1)}(t_2) =\frac{1}{2\pi}\frac{1}{4}\int_{-\pi}^{\pi} dk\,\bigl|\partial_{t_2} \hat{d}_{\text{SSH}}^{(1)}(k,t_2)\bigr|^2=\frac{1}{8\pi}\int_{-\pi}^{\pi} dk  \frac{t_1^2t_2^2\sin^4{k}}{(t_1^2 + t_2^2 -2t_1t_2\cos{k})^3}= \frac{3\left(t_1^2+t_2^2-\left|t_1^2-t_2^2\right|\right)^2}
{128\,\left|t_1^2-t_2^2\right|\,t_1^2 t_2^2}\,.
\end{align}
As a result, the first component of fidelity susceptibility for the SSH model is explicitly given by,
\begin{align}\label{eq:FS_SSH}
   \chi_F^{\text{SSH}(1)}(t_2) =
\begin{cases}
\dfrac{3t_2^2}{32\,t_1^2\left(t_1^2-t_2^2\right)}, & t_1>t_2, \\[1.2ex]
\dfrac{3t_1^2}{32\,t_2^2\left(t_2^2-t_1^2\right)}, & t_2>t_1\,.
\end{cases} 
\end{align}
Hence near the gap closing point ($t_1 = t_2$), the first component of the fidelity susceptibility diverges as,
\begin{align}
    \lim_{t_1 \to t_2} \chi_{F}^{\text{SSH}(1)}(t_2) \approx \frac{1}{|t_1-t_2|}\,.
\end{align}
Using Eq.~\eqref{eq:kry_sprd_ssh_full}, the Krylov spread complexity for the ground state of the SSH model as previously obtained in Appendix~\ref{appendix:b} and also see Eq.~\eqref{eq:kry_sprd_ssh_full} is given as follows,
\begin{align}\label{eq:Com_I1}
    C^{(1)}_{\text{SSH}}(t_1,t_2)
&=\frac12+\Re(\alpha^*\beta)\,\frac{\delta\,K(m)+s\,E(m)}{\pi t_1} = C_{\text{SSH}}(t_1,t_2)\,,
\end{align}
where we have,
\begin{align}
s=t_1+t_2\,,\qquad \delta=t_1-t_2\,,\qquad
m=\frac{4t_1t_2}{(t_1+t_2)^2}=\frac{4t_1t_2}{s^2}\,,
\qquad
1-m=\left(\frac{\delta}{s}\right)^2\,.
\end{align}
We emphasize that the in the spread complexity on the $x-$component is non-vanishing and hence we have introduced a superscript $1$. By using Eq.~\eqref{eq:I1} (see Appendix~\ref{appendix:b} for a detailed derivation), we obtain that the leading order singularity in the first derivative of $C^{(1)}_{\text{SSH}}(t_1,t_2)$ near the gap closing point is given by,
\begin{align}\label{eq:I1_div}
    \lim_{t_1 \to t_2}\partial_{t_2}C^{(1)}_{\text{SSH}}(t_1,t_2) \sim  \ln \frac{1}{|t_1-t_2|}\,.
\end{align}
Hence, to summarize, we have the following singular behavior near the gap closing point for the SSH model,
\begin{align}
    \lim_{t_1 \to t_2}\partial_{t_2} C_{\text{SSH}}(t_1,t_2) = \lim_{t_1 \to t_2}\partial_{t_2} C_{\text{SSH}}^{(1)}(t_1,t_2) \sim \ln \frac{1}{|t_1-t_2|} \,, \qquad \lim_{t_1 \to t_2} \sqrt{\chi_F^{\text{SSH}(1)}(t_2)} \sim \frac{1}{\sqrt{|t_1-t_2|}}\,,
\end{align}
As a result, like the massive Dirac Hamiltonian, we clearly see that the divergence in the fidelity susceptibility is power law, whereas for the derivative of the Krylov spread complexity it is logarithmic and hence the inequality holds, Eq.~\eqref{eq:fs_component} near the gap closing point $t_1=t_2$. Similar to the massive Dirac Hamiltonian case, we would also like to understand the the inequality Eq.~\eqref{eq:fs_component} much away from the gap closing point i.e., $t_2\gg t_{1}$. We define the ratio,
\begin{align}\label{eq:ratio_SSH}
   R(t_2) = \frac{|\partial_{t_2}C_{\text{SSH}}^{(1)}(t_1,t_2)|}{4\pi |\mathcal{Q}_{1}| \sqrt{\chi_F^{\text{SSH}(1)}(t_2)}}\,, \qquad  \mathcal{Q}_{1} = \frac{\Re(\alpha^*\beta)}{2\pi}\,. 
\end{align}
By using Eq.~\eqref{eq:FS_SSH}, for large $t_2$, we have the following,
\begin{align}\label{eq:den}
   \lim_{t_2\to\infty} 2\sqrt{\chi^{(1)}_{F,\mathrm{SSH}}(t_2)}
=
\frac{\sqrt{3}\,t_1}{2\sqrt{2}\,t_2^2}\,.
\end{align}
Similarly, we also have,
\begin{align}
    m \sim 4x - 8x^2 + O(x^3)\,, \qquad m = \frac{4x}{(1+x)^2}\,, \qquad x = \frac{t_1}{t_2}\to0\,.
\end{align}
Now using Eq.~\eqref{eq:exp_elliptic} for the series expansion of the elliptic integrals for small argument we have the following,
\begin{align}\label{eq:num}
   \lim_{t_2\to\infty} |\partial_{t_2}C_{\text{SSH}}^{(1)}(t_1,t_2)| = \lim_{t_2\to\infty}|\Re(\alpha^*\beta)\,\partial_{t_2}\mathcal{I}_1(t_1,t_2)| =|\Re(\alpha^*\beta)|\,\frac{t_1}{2t_2^2}\,.
\end{align}
Using Eq.~\eqref{eq:num} and Eq.~\eqref{eq:den} we have the following,
\begin{align}
    \lim_{t_2\to\infty} R(t_2) = \lim_{t_2\to\infty} \frac{|\partial_{t_2}C_{\text{SSH}}^{(1)}(t_1,t_2)|}{4\pi |\mathcal{Q}_{1}| \sqrt{\chi_F^{\text{SSH}(1)}(t_2)}} = \sqrt{\frac{2}{3}} < 1\,.
\end{align}
\subsection{Ratio of first derivative of Krylov complexity to the fidelity susceptibility }
In this subsection we are going to understand the reason for obtaining $R(\lambda) \to \sqrt{2/3}$ for massive Dirac Hamiltonian, cf. Eq.~\eqref{eq:ratio_MD} and also for SSH model, cf. Eq.~\eqref{eq:ratio_SSH}. Consider the following ratio,
\begin{align}\label{eq:ratio_general}
    \lim_{\lambda\to\infty} R(\lambda) = \lim_{\lambda\to\infty} \frac{|\partial_{\lambda}C^{(i)}(\lambda)|}{4\pi |\mathcal{Q}_{i}| \sqrt{\chi_F^{(i)}(\lambda)}} \,.
\end{align}
The parameter $\lambda$ can be tuned to go across the gap closing point. For instance, $\lambda=t_2$ for the SSH model and $\lambda = \mu$ for the massive Dirac Hamiltonian respectively. It is important to note that we are comparing only the non-vanishing components of the $\partial_{\lambda} C^{(i)}(\lambda)$. In the case of SSH model, it is the $x-$component and for the massive Dirac Hamiltonian, it the $z-$component respectively. In both these cases it can readily be seen that,
\begin{align}\label{eq:universal_condition}
  \lim_{\lambda\to\infty}  \partial_{\lambda} \hat{d}_i(k,\lambda)\sim f(\lambda)\sin^2{k}\,, 
\end{align}
where $f(\lambda)$ is a function of the parameter $\lambda$ that in general will be different for SSH model and the massive Dirac Hamiltonian. The important point is that this function $f(\lambda)$ gets canceled out in the ratio function, see Eq.~\eqref{eq:ratio_general} and hence we find,
\begin{align}\label{eq:ratio_general}
    \lim_{\lambda\to\infty} R(\lambda) = \frac{
\left|
\displaystyle \int_{0}^{\pi} dk\,
\sin^2{k}
\right|
}{
\sqrt{\pi}
\left[
\displaystyle \int_{0}^{\pi} dk\,
\left(\sin^2{k}\right)^2
\right]^{1/2} 
} = \sqrt{\frac{2}{3}}\,.
\end{align}
As a result for all Hamiltonians of the form $H(\lambda)=\pmb{\sigma} \cdot \textbf{d}(\lambda)$, which satisfy the condition Eq.~\eqref{eq:universal_condition}, will automatically satisfy Eq.~\eqref{eq:ratio_general}. This includes a plethora of models such as the SSH model, massive Dirac Hamiltonian, 1D-Kitaev chain and many more.

\section{Duality in the Su-Schrieffer-Heeger (SSH) model}\label{appendix:e}

In this section, we are going to present the details of the duality that exist in Su-Schrieffer-Heeger (SSH) model. Duality is between trivial and the topological phase of the SSH model. Let us start by writing the Hamiltonian for the SSH model, where we clearly see the duality,
\begin{align}
    H_{\text{I}}(k) &= (t - rt\cos\,k) \cdot \sigma_{x} + rt\sin\,k \cdot \sigma_{y} \,, \label{eq:ssh_ham_r1_appen_E} = \left[\begin{array}{cc}
0 & t-rt\,e^{ik}\\
t-rt\,e^{-ik} & 0
\end{array}\right]=\left[\begin{array}{cc}
0 & f_{\text{I}}(k)\\
f_{\text{I}}(k)^{*} & 0
\end{array}\right] \\
    H_{\text{II}}(k) &= \left(t - \frac{t}{r}\cos\,k\right) \cdot \sigma_{x} + \frac{t}{r}\sin\,k \cdot \sigma_{y} =\left[\begin{array}{cc}
0 & t-\frac{t}{r}\,e^{ik}\\
t-\frac{t}{r}\,e^{-ik} & 0
\end{array}\right]=\left[\begin{array}{cc}
0 & f_{\text{II}}(k)\\
f_{\text{II}}(k)^{*} & 0
\end{array}\right]\,,  \label{eq:ssh_ham_r2_appen_E}
\end{align}
where we have parametrized the SSH model Hamiltonian in terms of $r$ and $t$, in contrast to previous version of the Hamiltonian, cf. Eq.~\eqref{eq:ssh_ham_appen_a}. By setting $t=t_{1}$ and $r=t_{2}/t_{1}$, we would recover the SSH Hamiltonian we previously discussed, Eq.~\eqref{eq:ssh_ham_appen_a}. Let us start by calculating the winding number so that we determine the phase of each of the Hamiltonian $H_{\text{I}}(k)$, Eq.~\eqref{eq:ssh_ham_r1_appen_E} and $H_{\text{II}}(k)$, Eq.~\eqref{eq:ssh_ham_r2_appen_E}. Winding number is calculated as follows using Appendix~\ref{subsec:winding_ssh},
\begin{align}
\nu_{\text{I}} & =\frac{1}{2\pi i}\int_{-\pi}^{\pi}dk\partial_{k}\ln f_{\text{I}}(k) = \frac{1}{2\pi i}\oint_{|z|=1}dz\frac{1}{z-\frac{1}{r}} = \begin{cases}
    0\,,\qquad(\text{for }r < 1) \\
    1\,,\qquad(\text{for }r > 1)
\end{cases} \,, \\
\nu_{\text{II}} & =\frac{1}{2\pi i}\int_{-\pi}^{\pi}dk\partial_{k}\ln f_{\text{II}}(k) = \frac{1}{2\pi i}\oint_{|z|=1}dz\frac{1}{z-r} = \begin{cases}
    1\,,\qquad(\text{for }r < 1) \\
    0\,,\qquad(\text{for }r > 1)
\end{cases} \,,
\end{align}
hence, we find the duality between SSH Hamiltonians $H_{\text{I}}(k)$ and $H_{\text{II}}(k)$, cf.~\eqref{eq:ssh_ham_r1_appen_E} and Eq.~\eqref{eq:ssh_ham_r2_appen_E}, as a function of parameter $r$. If $r>1$ then, $H_{\text{I}}(k)$ is in topological phase and $H_{\text{II}}(k)$ in trivial phase, and if $r < 1$, then $H_{\text{I}}(k)$ is in trivial and $H_{\text{II}}(k)$ in topological phase. Next, we would like to emphasize that two Hamiltonians $H_{\text{I}}(k)$ and $H_{\text{II}}(k)$ are not unitarily related for $r\neq 1$. This can be seen by computing the determinant as follows,
\begin{align}
    \text{det} \big[ H_{\text{I}}(k) \big] &= t^{2} + r^{2}t^{2} - 2rt\cos k\,,\\
    \text{det} \big[ H_{\text{II}}(k) \big] &= t^{2} + \frac{t^{2}}{r^{2}} - \frac{2t}{r}\cos k\,,
\end{align}
and hence, there does not exist any unitary that transforms $H_{\text{I}}(k)$ to $H_{\text{II}}(k)$, and vice-versa.

\subsection{Duality in the fidelity susceptibility}
Next, let us move on to show how this duality manifests itself in fidelity susceptibility. Let us start with the following unitary transformation such that we have,
\begin{align}
    (\sigma_{x},\,\sigma_{y},\,\sigma_{z}) \to (\sigma_{x},\,\sigma_{z},\,-\sigma_{y})\,.
\end{align}
This implies, that we have the following,
\begin{align}
    H_{\text{I}}(k) = \textbf{d}_{\text{I}}(k)\cdot \pmb{\sigma}\, \qquad \text{and} \qquad H_{\text{II}}(k) = \textbf{d}_{\text{II}}(k)\cdot \pmb{\sigma}\,,
\end{align}
where,
\begin{align}
\textbf{d}_{\text{I}}(k) &=t(1 - r\,\cos k, \; 0,\; r\, \sin k)  =  t(\text{Re}[z_{\text{I}}], \; 0,\; \text{Im}[z_{\text{II}}] )\,,\\
\textbf{d}_{\text{II}}(k) &=t \left(1 - \frac{1}{r}\,\cos k, \; 0,\; \frac{1}{r}\, \sin k \right) = t(\text{Re}[z_{\text{II}}], \; 0,\; \text{Im}[z_{\text{II}}] )\,,
\end{align}
where we have defined the following,
\begin{align}
    z_{\text{I}} = 1-r\,e^{-ik}\,,\, \qquad z_{\text{II}} = 1-\frac{1}{r}\,e^{-ik}\,. 
\end{align}
Next, consider the following,
\begin{align}
    z_{\text{II}} = -\frac{e^{-ik}}{r}(1-r\,e^{ik}) = -\frac{e^{-ik}}{r}\cdot z_{\text{I}}^{*} \, \Rightarrow \text{Arg}(z_{\text{II}}) = (2n+1)\pi - k - \text{Arg}(z_{\text{I}})\,,
\end{align}
where $n$ is an integer. For simplicity in the notation, we shall denote,
\begin{align}
    \varphi_{\text{I}}(k,r) &= \text{Arg}(z_{\text{I}})\,,\qquad 
    \varphi_{\text{II}}(k,r) = \text{Arg}(z_{\text{II}}) = (2n+1)\pi - k - \varphi_{\text{I}}(k,r) \,.\label{eq:fidelity_duality_argument}
\end{align}
Inorder to obtain fidelity susceptibility, we start by obtaining $\hat{d}_{\text{I}}$ and $\hat{d}_{\text{II}}$,
\begin{align}
    \hat{d}_{\text{I}}(k,r) = \frac{\textbf{d}_{\text{I}}(k,r)}{|\textbf{d}_{\text{I}}(k,r)|} = \big[\cos \varphi_{\text{I}}(k,r),\; 0, \; \sin \varphi_{\text{I}}(k,r)\big]\,,\\
    \hat{d}_{\text{II}}(k,r) = \frac{\textbf{d}_{\text{II}}(k,r)}{|\textbf{d}_{\text{II}}(k,r)|} = \big[\cos \varphi_{\text{II}}(k,r),\; 0, \; \sin \varphi_{\text{II}}(k,r)\big]\,,
\end{align}
From here, we have,
\begin{align}
\partial_r \hat{d}_{\text{I}}(k,r) &= \big[ -\sin \varphi_{\text{I}}(k,r) \cdot \partial_r \varphi_{\text{I}}(k,r),\;0, \; \cos \varphi_{\text{I}}(k,r) \, \partial_r \varphi_{\text{I}}(k,r) \big] \, \Rightarrow \, 
|\partial_r \hat{d}_{\text{I}}(k,r)|^{2} = |\partial_r \phi_{I}(k,r)|^{2}\,.
\end{align}
Now recall the definition of fidelity susceptibility, Eq.~\eqref{eq:FS_final} and hence we have,
\begin{align}
    \chi_{F}^{(\text{I})}(k,r) &= \frac{1}{4} |\partial_r \hat{d}_{\text{I}}(k,r)|^{2} = \frac{1}{4} |\partial_r \varphi_{\text{I}}(k,r)|^{2} \,,\\
    \chi_{F}^{(\text{II})}(k,r)  &= \frac{1}{4} |\partial_r \varphi_{\text{II}}(k,r)|^{2} \,.
\end{align}
Let us consider a simple exercise, where we replace $r$ with the function of $r$, i.e., $f(r)$. This implies,
\begin{align}
    \partial_r \varphi_{\text{I}}\big[k,f(r)\big] = \partial_{f(r)} \varphi_{\text{I}}\big[k,f(r)\big] \cdot \partial_{r} f(r)\,,
\end{align}
and for $f(r) = 1/r$, we find,
\begin{align}
    \partial_r \varphi_{\text{I}}\big[k,r^{-1}\big] = -\frac{1}{r^{2}}\cdot \partial_{r^{-1}} \varphi_{\text{I}}\big[k,r^{-1}\big]\, \Rightarrow \left|\partial_r \varphi_{\text{I}}\big[k,r^{-1}\big]\right|^{2} = \frac{1}{r^{4}}\cdot \left|\partial_{r^{-1}} \varphi_{\text{I}}\big[k,r^{-1}\big]\right|^{2}\,,
\end{align}
By using Eq.~\eqref{eq:fidelity_duality_argument} and averaging over the Brillouin zone i.e., $k\in [-\pi,\pi]$, we find,
\begin{align}
    \chi_{F}^{(\text{II})}(r) = \frac{1}{r^{4}}\cdot \chi_{F}^{(\text{I})}\left( \frac{1}{r} \right)\,,
\end{align}
and hence, the duality between $H_{\text{I}}(k)$, Eq.~\eqref{eq:ssh_ham_r1_appen_E} and $H_{\text{II}}$, Eq.~\eqref{eq:ssh_ham_r2_appen_E} manifests itself also in the fidelity susceptibility.

\subsection{Duality in the Krylov spread complexity}
In this section, we try to show that this duality also manifests itself in the Krylov spread complexity. In order to see this, we first express the integral $\mathcal{I}_1$, in terms of the parameter $\textit{r}$, cf. Eq.~\eqref{eq:ssh_ham_r1_appen_E} and Eq.~\eqref{eq:ssh_ham_r2_appen_E}. Using Eq.~\eqref{eq:integral_I1}, we obtain the following,
\begin{align}
    \mathcal{I}_1(r) = \frac{(1-r)\,K(m)+(1+r)\,E(m)}{\pi} \,, \qquad  m=\frac{4r}{(1+r)^2}\,.
\end{align}
Hence, under the transformation, $r\mapsto \frac{1}{r}$ we have the following transformation rule for the integral $\mathcal{I}_1$,
\begin{align}\label{eq:duality_transformation_integrals}
\mathcal{I}_1(r) = r\,\mathcal{I}_1\!\left(\frac{1}{r}\right)+\frac{2(1-r)}{\pi}\,K(m)\,.
\end{align}
where we have used the fact that $m$ is invariant under the duality transformation $r \leftrightarrow \frac{1}{r}$. As a result, the Krylov spread complexity expressed as a function of parameter $r$ is given by,
\begin{align}\label{eq:Complexity_exp}
    C(r)=\frac{1}{2}+\Re(\alpha^*\beta)\,\mathcal{I}_1(r)\,.
\end{align}
Hence, using Eq.~\eqref{eq:duality_transformation_integrals} and Eq.~\eqref{eq:Complexity_exp}, we find that the Krylov spread complexity transforms in the following manner under the duality transformation $r\mapsto\frac{1}{r}$,
\begin{align}\label{eq:dual_complexity}
    C\!\left(\frac{1}{r}\right)&=\frac{1}{r}\bigl[C(r)-\mathcal{H}(r)\bigr],
\qquad
\mathcal{H}(r)=\frac{1-r}{2}+\frac{2\Re(\alpha^*\beta)(1-r)}{\pi}K(m).
\end{align}
Thus, knowing the Krylov spread complexity in one phase completely determines its value in the dual phase and $C(r=1)$ is mapped to the same point under the map $r\leftrightarrow \frac{1}{r}$. Next, differentiating Eq.~\eqref{eq:dual_complexity} with respect to $r$, we obtain the following,
\begin{align}
    C'\!\left(\frac{1}{r}\right)=C(r)-\mathcal{H}(r)-r\,C'(r)+r\,\mathcal{H}'(r).
\end{align}
Now, let us examine this relation at the self dual point of the map which is at $r=1$. Using the fact that $\mathcal{H}(r=1)=0$, we obtain the following,
\begin{align}\label{eq:constraint}
  2C'(1)=C(1)+\mathcal{H}'(1).  
\end{align}
We note that Eq.~\eqref{eq:constraint} is a constraint on the first derivative of the Krylov spread complexity at the self-dual point, $r=1$. We would expect the left hand side of Eq.~\eqref{eq:constraint} to diverge logarithmically at the self dual point $r=1$ for all reference states such that $\Re(\alpha^*\beta)\neq0$ [see Eq.~\eqref{eq:I1_div}], this is a check with our analytical results. From Eq.~\eqref{eq:constraint}, it is clear that the divergence of $C'(1)$ is solely controlled by the $\mathcal{H}'(1)$ term because there is no divergence in Krylov spread complexity $C(r)$ for all $r$. Differentiating the function $\mathcal{H}_1(r)$ gives the following,
\begin{align}
    \mathcal{H}'(r)
=
-\frac{1}{2}
+\frac{2\Re(\alpha^*\beta)}{\pi}\left[-K(m)+(1-r)\frac{dK}{dm}\,m'(r)\right].
\end{align}
As a result, near $r=1$, we have the leading singular behavior,
\begin{align}
   \lim_{r \to 1} \mathcal{H}'(r)\sim -\frac{2a}{\pi}K(m)
\sim -\frac{2a}{\pi}\ln\!\left(\frac{1}{|1-r|}\right)\,, \qquad m=\frac{4r}{(1+r)^{2}}\to 1\,.
\end{align}
Hence, the leading order singularity in the Krylov spread complexity is obtained as follows,
\begin{align}
    C'(r)\sim \frac{\Re(\alpha^*\beta)}{\pi}\ln|1-r|
\qquad (r\to 1).
\end{align}
 Thus, the first derivative of the integrated Krylov complexity density $C(r)$ diverges logarithmically (for all reference states with $\Re(\alpha^*\beta)\neq0$) at the self-dual point $r=1$, in agreement with the previous analytical results, cf. Appendix~\ref{appendix:b}. We would like to emphasize that the duality relation Eq.~\eqref{eq:dual_complexity} also holds even when one considers implicit momentum \textit{k-dependent} reference states in Eq.~\eqref{eq:ref_class_1}. In this case, both the integrals $\mathcal{I}_1$ and $\mathcal{I}_3$ contribute to the expression of the integrated Krylov complexity. It turns out that the graph of the ratio $R$ for the $\text{SSH}$ model [see Eq.~\eqref{eq:ratio_SSH}] plotted as a function of $\log(t_2/t_1)$ is symmetric about $t_1=t_2$ (see Fig.~\ref{fig:rlambda}). This is a clear manifestation of the duality in the $\text{SSH}$ model. Note that the ratio $R$ can be expressed as following,
\begin{align}
R(r)
=
\frac{
\displaystyle \Bigg|\int_{0}^{\pi} dk \,
\frac{\sin^{2} k}{D_r^{3}(k)}
\Bigg|}{
\displaystyle
\sqrt{\pi}
\left[
\int_{0}^{\pi} dk \,
\left(
\frac{\sin^{2} k}{D_r^{3}(k)}
\right)^2
\right]^{1/2}
}\,, \qquad D_r(k)=\sqrt{1+r^2-2r\cos k}\,, \qquad r=\frac{t_2}{t_1}\,.
\end{align}
Now, under the duality transformation, $r \to \frac{1}{r}$, we have the following,
\begin{align}\label{eq:D_duality}
   D_{1/r}(k)
=
\sqrt{1+\frac{1}{r^2}-\frac{2}{r}\cos k}
=
\frac{1}{r}\sqrt{1+r^2-2r\cos k}
=
\frac{D_r(k)}{r}. 
\end{align}
Hence, using Eq.~\eqref{eq:D_duality}, it is clear that both numerator and denominator scale with an overall multiplicative factor of $r^3$, which gets canceled in the ratio $R(r)$. Hence, we have the following,
\begin{align}\label{eq:frac_r_duality}
    R\Big(\frac{1}{r}\Big) = R(r)\,.
\end{align}
As a result, when plotted in terms of the variable $u=\log(r)$, the duality transformation $r \leftrightarrow \frac{1}{r}$ becomes equivalent to $u \leftrightarrow -u$. Finally, Eq.~\eqref{eq:frac_r_duality} becomes equivalent to,
\begin{align}
    R(u) = R(-u)\,, \qquad u =\log (r)\,.
\end{align}
Hence, $R(u)$ is an even function of $u$, and is symmetric about $u=0$, which corresponds to $r=1$.

\begin{figure}[htbp]
        \centering
        \includegraphics[width=0.5\linewidth]{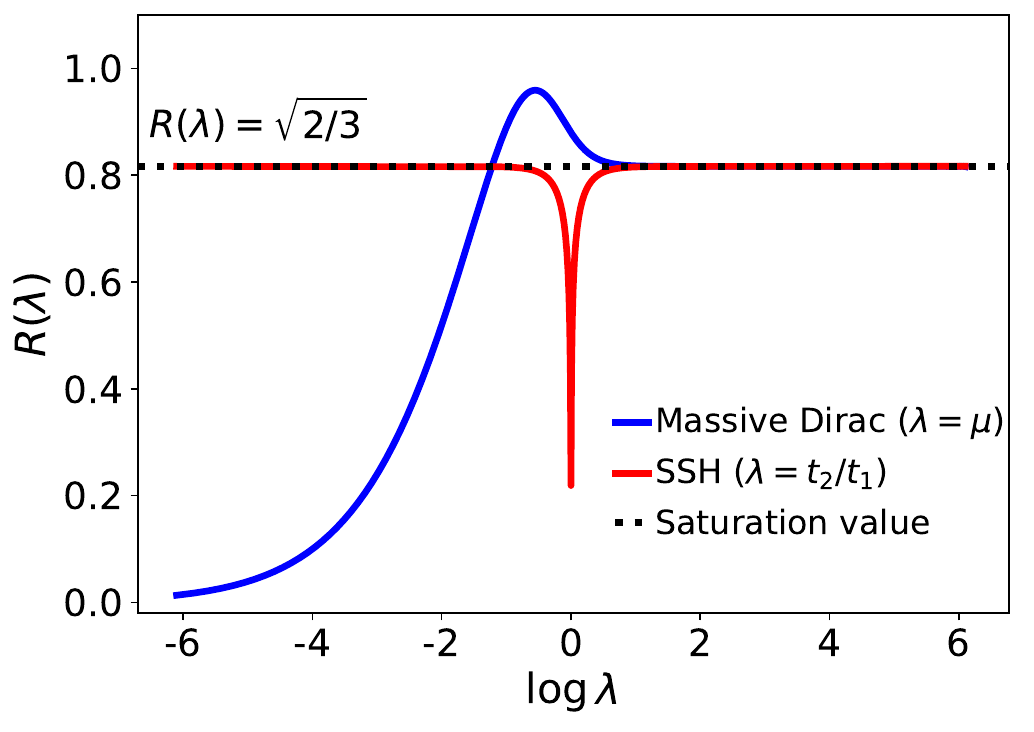}
        \caption{Ratio of nonzero component of derivative of spread complexity to corresponding component of fidelity susceptibility for the massive Dirac Hamiltonian and the SSH model. For the massive Dirac Hamiltonian, $\lambda = \mu$, whereas for SSH, $\lambda = r = t_{2}/t_{1}$. The minimum of the ratio for SSH occurs at $r=1$. At large $\lambda$, the ratio $R(\lambda)$ saturates to a value of $\sqrt{2/3}$.}
        \label{fig:rlambda}
\end{figure}

\section{Krylov spread complexity in the non-Hermitian SSH model}\label{appendix:f}

In this section, we show that the Krylov spread complexity also captures the topological phase transition in the non-Hermitian SSH model under periodic boundary conditions. We work with the non-Hermitian SSH model discussed in~\cite{PhysRevLett.121.086803},
\begin{equation}
    H = \sum_j \left [ \left (t_1+\frac{\gamma}{2} \right ) c_{A,j}^{\dagger} c_{B,j}  + \left (t_1-\frac{\gamma}{2} \right ) c_{B,j}^{\dagger} c_{A,j}  + \left ( t_2 c_{A,j+1}^{\dagger}c_{B,j} + h.c. \right ) \right ].
\end{equation}
Under periodic boundary conditions, this model has topological phase transitions at $t_2 = \pm \left ( t_1 \pm \frac{\gamma}{2} \right )$. After performing a Fourier transform and making the unitary transformation of the basis $(\sigma_x, \sigma_y, \sigma_z) \to (\sigma_x, \sigma_z, -\sigma_y)$, the Hamiltonian is written as,
\begin{equation}
    H = \sum_k \begin{pmatrix}
        c_k^{\dagger} & d_k^{\dagger} 
    \end{pmatrix} h(k) \begin{pmatrix}
        c_k \\ d_k
    \end{pmatrix} \quad ; \quad h(k) = \begin{pmatrix}
        R_3 & R_1 \\ R_1 & -R_3
    \end{pmatrix},
\end{equation}
where $R_1 = t_1 - t_2 \cos k$ and $R_3 = t_2 \sin k + \frac{i\gamma}{2}$. The eigenvalues of each $h(k)$ are $R_{\pm}$, where,
\begin{equation}
    R^2 = R_1^2 + R_3^2 = t_1^2 + t_2^2 -2t_1 t_2 \cos k -\frac{\gamma^2}{4} + i\gamma t_2 \sin k.
\end{equation}
The ground state is the eigenvector which has $\text{Re}(E) < 0$ for all $k$. Since the Hamiltonian is now non-Hermitian, we must consider both the left and the right eigenvectors while constructing the Krylov basis. The unnormalized right and left eigenvectors for the ground state, $| \psi_g^R\rangle$ and $\langle \psi_g^L|$, are respectively defined using,
\begin{equation}
    h(k) |\psi_g^R\rangle = -R |\psi_g^R\rangle \quad ; \quad  \langle \psi_g^L| h(k) = -R\langle \psi_g^L|.
\end{equation}
Biorthogonal normalization $\langle \psi_g^L | \psi_g^R\rangle = 1$ gives us,
\begin{equation}
    |\psi_g^R\rangle = \frac{1}{\sqrt{R_1^2 + (R+R_3)^2}}\begin{pmatrix}
        R_1 \\ -(R_3 + R)
    \end{pmatrix} \quad ; \quad \langle \psi_g^L| = \frac{1}{\sqrt{R_1^2 + (R+R_3)^2}} \begin{pmatrix}
        R_1 & -(R_3 + R)
    \end{pmatrix}\,.
\end{equation}
\begin{figure}[t!]
    \centering
    \includegraphics[width=0.5\linewidth]{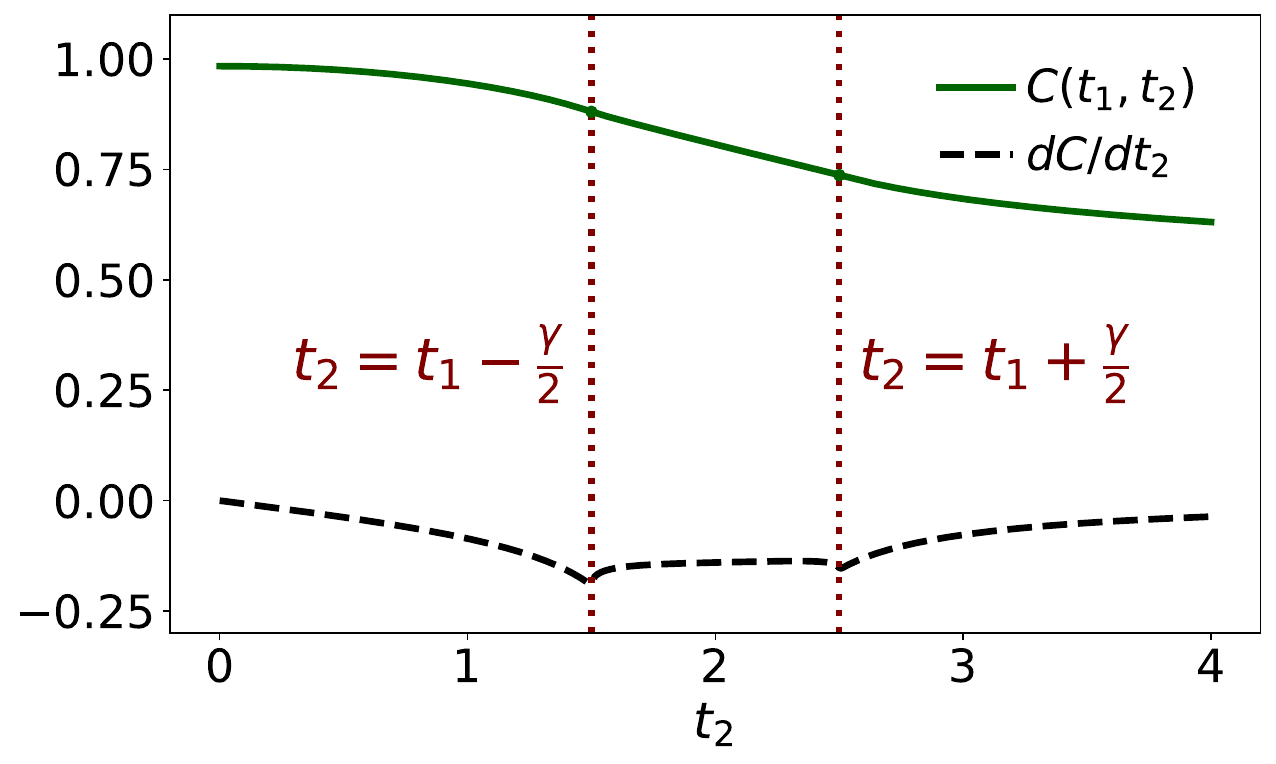}
    \caption{Krylov spread complexity for the non-Hermitian SSH model with $t_1 = 2$ and $\gamma = 1$ under periodic boundary conditions (PBC). The reference state is chosen using $\alpha=\beta=1/2$ in Eq.~(\ref{eq:right_ref}) and Eq.~(\ref{eq:left_ref}). The non-analytic behaviour of the complexity $C(t_1,t_2)$ occurs at the PBC gap closing points, $t_2 = t_1 \pm \frac{\gamma}{2}$. This is illustrated using the cusps in $dC(t_1,t_2)/dt_2$.}
    \label{fig:nh_ssh}
\end{figure}
It is important to note that the normalization factor is a complex number. The right reference state is once again taken to be a generic state on the Bloch sphere,
\begin{equation}\label{eq:right_ref}
    |\psi_0^R\rangle = \alpha |\uparrow\rangle + \beta |\downarrow\rangle, \text{ } |\alpha|^2 + |\beta|^2=1.
\end{equation}
The left reference state is now fixed by biorthonormality. Imposing the condition $\langle \psi_0^L | \psi_0^R \rangle = 1$, we find,
\begin{equation}\label{eq:left_ref}
    \langle \psi_0^L | = \alpha^* \langle \uparrow | + \beta^* \langle \downarrow |\,.
\end{equation}
We now employ a bi-Lanczos algorithm, which is a generalization of the Lanczos algorithm for a non-Hermitian circuit Hamiltonian. Denoting the circuit Hamiltonian as $H_c$, we can construct the right Krylov subspace,
\begin{equation}
    \mathbb{K}_2^R = \text{span}\{|\psi_0^R\rangle, H_c |\psi_0^R\rangle \},
\end{equation}
and the left Krylov subspace,
\begin{equation}
    \mathbb{K}_2^L = \text{span} \{ \langle \psi_0^L |, \langle \psi_0^L | H_c \}.
\end{equation}
Denoting the right Krylov basis vectors as $|\mathcal{K}_0^R \rangle$ and $|\mathcal{K}_1^R \rangle$ and the left Krylov basis vectors as $\langle \mathcal{K}_0^L|$ and $\langle \mathcal{K}_1^L|$ (where $|\mathcal{K}_0^R \rangle = |\psi_0^R \rangle$ and $\langle \mathcal{K}_0^L | = \langle \psi_0^L |$), we have the biorthonormality condition $\langle \mathcal{K}_i^L | \mathcal{K}_j^R \rangle = \delta_{ij}$. The spread complexity for a given momentum mode is once again determined by the overlaps of the Krylov vectors with the target state. We use the prescription,
\begin{equation}
    C_k = \frac{|w_1|}{|w_0| + |w_1|},
\end{equation}
where,
\begin{equation}
    w_0 = \langle \mathcal{K}_0^L | \psi_g^R \rangle \langle \psi_g^L | \mathcal{K}_0^R \rangle, \text{ } w_1 = \langle \mathcal{K}_1^L | \psi_g^R \rangle \langle \psi_g^L | \mathcal{K}_1^R \rangle.
\end{equation}
From Eq.~\eqref{eq:right_ref} and Eq.~\eqref{eq:left_ref}, we have,
\begin{equation}
    |\mathcal{K}_0^R \rangle = \begin{pmatrix}
        \alpha \\ \beta
    \end{pmatrix}, \text{ } \langle \mathcal{K}_0^L | = \begin{pmatrix}
        \alpha^* & \beta^*
    \end{pmatrix},
\end{equation}
where we are working in the $\{ |\uparrow \rangle, |\downarrow \rangle \}$ basis. From the conditions $\langle \mathcal{K}_1^L | \mathcal{K}_0^R \rangle = 0$ and $\langle \mathcal{K}_0^L | \mathcal{K}_1^R \rangle = 0$, we have,
\begin{equation}
    |\mathcal{K}_1^R \rangle = \begin{pmatrix}
        \beta^* \\ -\alpha^*
    \end{pmatrix}, \text{ } \langle \mathcal{K}_1^L | = \begin{pmatrix}
        \beta & -\alpha 
    \end{pmatrix}.
\end{equation}
We now have the following,
\begin{equation}
\begin{split}
    & w_0 = \frac{\left ( \alpha R_1 - \beta (R + R_3) \right ) \left ( \alpha^* R_1 - \beta^* (R + R_3) \right )}{R_1^2 + (R+R_3)^2}, \\
    & w_1 = \frac{\left ( \alpha (R + R_3) + \beta R_1 \right ) \left ( \alpha^* (R + R_3) + \beta^* R_1 \right )}{R_1^2 + (R + R_3)^2}.
\end{split}
\end{equation} 
The complexity per momentum mode $C_k$ thus has a complicated form, and integrating it over the Brillouin zone must be done numerically. Fig.~\ref{fig:nh_ssh} shows the complexity of the ground state of the non-Hermitian SSH model, and we see that non-analyticities in the complexity occur at precisely the gap closing points under periodic boundary conditions (PBC). We must emphasize that non-Hermitian Hamiltonians are highly sensitive to boundary conditions, and the topological phase transition point of the non-Hermitian SSH model under open boundary conditions (OBC) is famously known to be different from the PBC case~\cite{PhysRevLett.121.086803}. Detecting the OBC phase transition point using Krylov spread complexity requires us to know the exact circuit Hamiltonian as well as sophisticated numerical techniques. It is an interesting direction for future research.

\end{widetext}

\end{appendix}

\bibliography{references}

\end{document}